\documentclass[conference]{IEEEtran}
\IEEEoverridecommandlockouts

\usepackage{cite}
\usepackage{amsmath,amssymb,amsfonts}
\usepackage{graphicx}
\usepackage{textcomp}
\usepackage{xcolor}
\usepackage[hyphens]{url}

\usepackage{xspace}
\usepackage{bm}
\usepackage{stfloats}
\usepackage{kotex}
\usepackage{lipsum}
\usepackage{enumitem}
\usepackage{fancyhdr}
\usepackage{comment}
\usepackage{makecell}
\usepackage{tabularx}
\usepackage{threeparttable}
\usepackage{multirow}
\usepackage{comment}
\usepackage{xfrac}
\usepackage{multirow}
\usepackage{tcolorbox}
\usepackage{booktabs}
\usepackage{color,soul}
\usepackage[utf8]{inputenc}
\usepackage[english]{babel}
\usepackage{balance}
\usepackage{circledsteps}
\usepackage{calc}
\usepackage{adjustbox}
\usepackage{tikz}
\usepackage{textcase}
\usepackage{algorithm}
\usepackage{algpseudocode}
\usepackage[absolute,showboxes]{textpos}

\newcommand{\actcmd}{\texttt{ACT}\xspace}
\newcommand{\precmd}{\texttt{PRE}\xspace}
\newcommand{\rdcmd}{\texttt{RD}\xspace}
\newcommand{\wrcmd}{\texttt{WR}\xspace}
\newcommand{\refcmd}{\texttt{REF}\xspace}
\newcommand{\refcmds}{\texttt{REF}s\xspace}
\newcommand{\rfmcmd}{\texttt{RFM}\xspace}
\newcommand{\rfmcmds}{\texttt{RFM}s\xspace}
\newcommand{\refabcmd}{\texttt{REF}\textsubscript{ab}\xspace}
\newcommand{\refsbcmd}{\texttt{REF}\textsubscript{sb}\xspace}
\newcommand{\trc}{$t_{\mathrm{RC}}$\xspace}
\newcommand{\trcd}{$t_{\mathrm{RCD}}$\xspace}
\newcommand{\tras}{$t_{\mathrm{RAS}}$\xspace}
\newcommand{\trp}{$t_{\mathrm{RP}}$\xspace}
\newcommand{\twr}{$t_{\mathrm{WR}}$\xspace}
\newcommand{\trtp}{$t_{\mathrm{RTP}}$\xspace}

\newcommand{\trefw}{$t_{\mathrm{REFW}}$\xspace}
\newcommand{\trefi}{$t_{\mathrm{REFI}}$\xspace}
\newcommand{\trfc}{$t_{\mathrm{RFC}}$\xspace}

\newcommand{\trcdcsa}{$t_{\mathrm{RCD_{CSA}}}$\xspace}
\newcommand{\trascsa}{$t_{\mathrm{RAS_{CSA}}}$\xspace}
\newcommand{\tup}{$t_{\mathrm{UP}}$\xspace}
\newcommand{\twrcsa}{$t_{\mathrm{WR_{CSA}}}$\xspace}
\newcommand{\trpcsa}{$t_{\mathrm{RP_{CSA}}}$\xspace}

\newcommand{\eg}{\emph{e.g.}\xspace}
\newcommand{\ie}{\emph{i.e.}\xspace}

\newcommand{\rh}{RH\xspace}
\newcommand{\nbo}{\texttt{N}\textsubscript{BO}\xspace}
\newcommand{\nrhagg}{\texttt{N}\textsubscript{RH}\xspace}
\newcommand{\nrhvic}{\texttt{N}\textsubscript{RHv}\xspace}
\newcommand{\nmit}{\texttt{N}\textsubscript{Mit}\xspace}
\newcommand{\taborec}{$t\mathrm{ABO_{Recovery}}$\xspace}
\newcommand{\taboact}{$t\mathrm{ABO_{ACT}}$\xspace}
\newcommand{\aboact}{\texttt{ABO}\textsubscript{ACT}\xspace}
\newcommand{\abodel}{\texttt{ABO}\textsubscript{Delay}\xspace}
\newcommand{\alert}{\texttt{Alert}\xspace}
\newcommand{\alerts}{\texttt{Alert}s\xspace}
\newcommand{\br}{\texttt{BR}\xspace}

\definecolor{revcolor}{RGB}{2, 107, 213}

\definecolor{lightergray}{gray}{0.95} 

\newcommand*{\Scale}[2][4]{\scalebox{#1}{$#2$}}

\newcommand{\blackcircled}[1]{%
    \tikz[baseline=(char.base)]{
        \node[shape=circle, draw, fill=black, inner sep=1pt] (char) {\textcolor{white}{\small #1}};
    }%
}

\def\BibTeX{{\rm B\kern-.05em{\sc i\kern-.025em b}\kern-.08em
    T\kern-.1667em\lower.7ex\hbox{E}\kern-.125emX}}

\TPMargin{5pt}

\newcommand{\copyrightstatement}{
    \begin{textblock}{0.84}(0.08,0.93)    
         \noindent
         \footnotesize
         \textcopyright\ 2026 IEEE. Personal use of this material is permitted. Permission from IEEE must be obtained for all other uses, in any current or future media, including reprinting/republishing this material for advertising or promotional purposes, creating new collective works, for resale or redistribution to servers or lists, or reuse of any copyrighted component of this work in other works.
    \end{textblock}
}
\begin{document}
\copyrightstatement

\title{PVAC: A RowHammer Mitigation Architecture Exploiting Per-victim-row Counting
}

\newcommand\iscaauthors{Jumin Kim\IEEEauthorrefmark{2}$^{*}$, Seungmin Baek\IEEEauthorrefmark{2}$^{*}$, Hwayong Nam\IEEEauthorrefmark{2}, Minbok Wi\IEEEauthorrefmark{3}, Nam Sung Kim\IEEEauthorrefmark{4}, Jung Ho Ahn\IEEEauthorrefmark{2}}

\newcommand\iscaaffiliation{\IEEEauthorrefmark{2}Seoul National University, \IEEEauthorrefmark{3}Samsung Electronics, \IEEEauthorrefmark{4}University of Illinois Urbana-Champaign}

\newcommand\iscaemail{\IEEEauthorrefmark{2}\{tkfkaskan1, qortmdalss, nhy4916, gajh\}@snu.ac.kr, \IEEEauthorrefmark{3}minbok.wi@samsung.com, \IEEEauthorrefmark{4}nskim@illinois.edu}

\author{
    \IEEEauthorblockN{\iscaauthors}
    \vspace{1.5mm} 
    \IEEEauthorblockA{\textit{\iscaaffiliation} \\ \iscaemail}
    \thanks{$^*$These authors contributed equally to this work.}
}

\maketitle

\thispagestyle{empty}
\pagestyle{plain}

\begin{abstract}

As DRAM scaling exacerbates RowHammer, DDR5 introduces per-row activation counting (PRAC) to track aggressor activity.
However, PRAC indiscriminately increments counters on every activation---including benign refreshes---while relying solely on explicit \rfmcmd operations for resets.
Consequently, counters saturate even in an idle bank, triggering cascading mitigations and degrading performance.
This vulnerability arises from a fundamental mismatch: PRAC tracks the \emph{aggressor} but aims to protect the \emph{victim}.

We present \textbf{P}er-\textbf{V}ictim-row h\textbf{A}mmered \textbf{C}ounting (PVAC), a victim-based counting mechanism that aligns the counter semantics with the physical disturbance mechanism of RowHammer. 
PVAC increments the counters of victim rows, resets the activated row, and naturally bounds counter values under normal refresh. 
To enable efficient victim-based updates, PVAC employs a dedicated counter subarray (CSA) that performs all counter resets and increments concurrently with normal accesses, without timing overhead. 
We further devise an energy-efficient CSA layout that minimizes refresh-induced counter accesses.
Through victim-based counting, PVAC supports higher hammering tolerance than PRAC while maintaining the same worst-case safety guarantee.
Across benign workloads and adversarial attack patterns, PVAC avoids spurious \alerts, eliminates PRAC timing penalties, and achieves higher performance and lower energy consumption than prior PRAC-based defenses.

\end{abstract}


\section{Introduction}
\label{sec:1_introduction}

Bits stored in a DRAM row (victim row) can flip if the rows physically adjacent to the victim row (referred to as aggressor rows) go through cycles of activation-precharge repeatedly, a phenomenon referred to as RowHammer (\rh)~\cite{isca-2014-para}.
From the perspective of an aggressor row, if the number of activation-precharge cycles does not surpass a threshold (\nrhagg), no bitflips occur in the victim rows.
Thus, by counting these cycles and refreshing the victim rows before the count reaches \nrhagg, we can prevent \rh.
Over the past decade, numerous hardware and software defense mechanisms~\cite{micro-2020-graphene, security-2024-abacus, hpca-2024-comet, isca-2022-hydra, isca-2019-twice, hpac-2024-start, isca-2018-CBT, DRAMSec-2021-panopticon, dac-2017-prohit, dac-2019-mrloc, arXiv-2024-proteas, arxiv-2025-marc, hpca-2021-blockhammer, micro-2024-breakhammer, hpca-2023-srs,asplos-2022-rrs, snp-2026-rowhammer, hpca-2025-chronus, hpca-2025-qprac, isca-2025-mopac, asplos-2025-moat, hpca-2022-mithril, micro-2023-cube, arxiv-2025-practical, MICRO-2024-mint, isca-2024-pride, isca-2025-dream, asplos-2024-rubix, hpca-2025-dapper, hpca-2025-autorfm, salt-2026-HPCA, mirza-2026-HPCA, CHaRM-2025-CCS} have been proposed to safeguard against a wide range of \rh attacks~\cite{dimva-2016-rowhammerjs, sec-2021-smash,security-2024-tossing,security-2024-yes,sp-2022-deepsteal,sec-2019-terminalbrain,atc-2018-throwhammer,blackhat-2015-privilege,ccs-2016-drammer,sec-2022-halfdouble,sp-2018-anotherflip,micro-2020-pthammer,sec-2016-onebitonecloud,sosp-2017-sgxbomb,sec-2024-gadgethammer,sp-2020-rambleed,sec-2020-deephammer,sp-2022-specHammer,sec-2016-fengshui,asplos-2025-marionette, sec-2024-sledgehammer, micro-2025-phammer, security-2025-gpuhammer}.

Per-row activation counting (PRAC) was recently introduced in the latest DDR5 JEDEC specification, which dedicates a small number of counter bits to each DRAM row to precisely count every activation~\cite{jedec-2024-ddr5}.
When a row is activated, PRAC increments its corresponding counter; these activations include activate (\actcmd), refresh (\refcmd), and refresh management (\rfmcmd) operations.
If the counter value exceeds a predefined threshold (\nbo), PRAC triggers an alert back-off (ABO) protocol that instructs the memory controller (MC) to issue an \rfmcmd command.
The \rfmcmd command performs preventive refreshes on its neighboring victim rows, which activate them, and also resets the counter value for the aggressor row.
We provide the pseudo code of PRAC in Algorithm~\ref{alg:prac}.

\begin{algorithm}[tb!]
\caption{An aggressor-based counting mechanism}
\label{alg:prac}
\small
\begin{algorithmic}[1]
\State \textbf{State:} per-row counter array $\mathit{cnt}[]$, back-off threshold \nbo, and \rfmcmd target queue $Q$
\vspace{0.2cm}
\State \textbf{procedure} \textsc{ACT}($r$):
\State \hspace{0.5cm} $\mathit{cnt}[r] \gets \mathit{cnt}[r] + 1$
\State \hspace{0.5cm} \textbf{if} $\mathit{cnt}[r] \ge $ \nbo \textbf{then}
\State \hspace{1.0cm} \textsc{Enqueue}$(Q, r)$
\State \hspace{1.0cm} \textsc{RFM}()\hfill // \textsc{Alert}
\vspace{0.2cm}
\State \textbf{procedure} \textsc{RFM}(): \hfill
\State \hspace{0.5cm} $r \gets \textsc{Dequeue}(Q)$
\State \hspace{0.5cm} \textbf{for each} row $v$ \textbf{in} $\mathit{Victims}(r)$ \textbf{do} 
\State \hspace{1.0cm} \textsc{ACT}($v$) \hfill // \rfmcmd-induced \actcmd
\State \hspace{0.5cm} $\mathit{cnt}[r] \gets 0$
\end{algorithmic}
\end{algorithm}

\begin{algorithm}[tb!]
\caption{A victim-based counting mechanism}
\label{alg:pvac}
\small
\begin{algorithmic}[1]
\State \textbf{State:} per-row counter array $\mathit{cnt}[]$, back-off threshold \nbo, and \rfmcmd target queue $Q$
\vspace{0.2cm}
\State \textbf{procedure} \textsc{ACT}($r$):
\State \hspace{0.5cm} $\mathit{cnt}[r] \gets 0$
\State \hspace{0.5cm} \textbf{for each} row $v$ \textbf{in} $\mathit{Victims}(r)$ \textbf{do} 
\State \hspace{1.0cm} $\mathit{cnt}[v] \gets \mathit{cnt}[v] + 1$
\State \hspace{1.0cm} \textbf{if} $\mathit{cnt}[v] \ge $ \nbo \textbf{then}
\State \hspace{1.5cm} \textsc{Enqueue}$(Q, v)$
\State \hspace{1.5cm} \textsc{RFM}() \hfill // \textsc{Alert}
\vspace{0.2cm}
\State \textbf{procedure} \textsc{RFM}():
\State \hspace{0.5cm} $r \gets \textsc{Dequeue}(Q)$
\State \hspace{0.5cm} \textsc{ACT}($r$) \hfill // \rfmcmd-induced \actcmd
\end{algorithmic}
\end{algorithm}

\begin{figure}[!tb]
  \center
  \includegraphics[width=0.86\columnwidth]{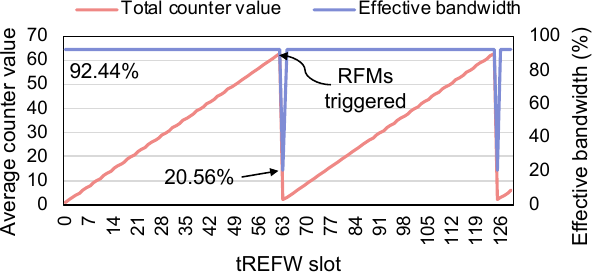}
  \vspace{-0.07in}
  \caption{Counter value per row and available bandwidth across the tREFW window, solely driven by normal refresh under PRAC when \nbo, \nmit, and blast radius are 64, 4, and 2, respectively.}
  
  \label{fig:domino}
  \vspace{-0.1in}
\end{figure}

However, PRAC's aggressor-based counting creates a structural asymmetry: counters increment on every activation (including periodic \refcmds) but reset only during explicit \rfmcmd commands~\cite{jedec-2024-ddr5}.
Because refreshes occur continuously, counter values monotonically increase even in an idle bank.
Consequently, as counters approach the threshold (\nbo), a single activation can trigger a chain reaction of \rfmcmd commands across the bank.
This domino effect severely degrades performance and can intermittently halt memory operations, causing denial-of-service (DoS) behavior even without malicious accesses.
Figure~\ref{fig:domino} illustrates the average counter value and the resulting available memory bandwidth of a DDR5 single-rank DIMM in the absence of memory accesses, configured with \nbo = 64, \nmit = 4, and a blast radius of 2, consistent with \nbo values evaluated in recent PRAC-based studies~\cite{hpca-2025-qprac, hpca-2025-chronus, isca-2025-mopac, asplos-2025-moat}.
Even when a bank is idle, the effective bandwidth is 92.44\% due to periodic refresh operations.
As \refcmd commands are issued continuously, the model shows that counters increase linearly.
At the beginning of the 63rd \trefw, the counter values of the earliest refreshed rows reach the \nbo threshold (\ie, 64), triggering a sequence of \rfmcmd commands.
As a result, the effective bandwidth collapses to 20.56\% within the 63rd \trefw, leading to intermittent but severe DoS behavior.
This issue becomes even worse when normal accesses are present, as their activations accelerate counter accumulation.

PRAC's counter organization further exacerbates the problem.
Prior in-DRAM mitigation schemes maintain counters in SRAM/CAM tables, enabling all counters to be reset in parallel once per \trefw~\cite{micro-2020-graphene, isca-2019-twice, isca-2018-CBT}.
However, as PRAC stores counter bits directly inside DRAM rows~\cite{jedec-2024-ddr5}, such parallel reset operations incur significant overhead.
This design forces counter accesses to incur an \actcmd-\precmd sequence, resulting in long latency.
Overall, PRAC-based mechanisms are fundamentally constrained not only by the one-directional counter increments but also by the high cost of counter resets, highlighting the need for a new counting organization.

To address this problem, we propose \emph{PVAC, \textbf{P}er-\textbf{V}ictim-row h\textbf{A}mmered \textbf{C}ounting}, which fundamentally shifts counting from aggressor rows to victim rows.
Similar to PRAC, each DRAM row has its dedicated counter.
However, instead of incrementing the counter of the row being activated (an aggressor row), we increment the counters of its victim rows. 
For simplicity,
we initially assume a blast radius of one (\ie, an aggressor row influences one row on each side).
If counter values are stored in an array $\mathit{cnt}[\,]$ and the index of the activated row is $r$, we reset $\mathit{cnt}[r]$ to $0$ and increment the counters of the potential victim rows ($\mathit{cnt}[r\pm1]$).
In PVAC, therefore, $\mathit{cnt}[r]$ indicates the vulnerability of the row itself against \rh, whereas in PRAC it indicates the vulnerability of the neighboring rows  (Algorithm~\ref{alg:pvac}).

Because every activation naturally resets the counter of the accessed row, PVAC prevents unnecessary accumulation.
While an \refcmd operation implicitly activates each row once per \trefw, PVAC resets the counter of the refreshed row during this activation.
Although \refcmd-induced activations may briefly increase a victim's counter, a subsequent \refcmd operation soon resets its counter back to zero, preventing long-term accumulation.
Even when normal accesses are interleaved, PVAC maintains this bounded behavior because each \actcmd and \refcmd also resets the activated row’s counter.

As PVAC updates the counter values of multiple rows on each activation, we employ a counter subarray (CSA)~\cite{DRAMSec-2021-panopticon,hpca-2025-chronus}.
The CSA does not share the data path with the data subarray (DSA), allowing counter updates for both aggressor and victim rows to proceed in parallel with normal memory accesses.
However, resetting counters during refresh could require activating multiple CSA rows.
Accordingly, we introduce an energy-efficient CSA organization that enables these updates with a single CSA activation.
This structure removes the overhead of multiple activations while preserving both performance and energy efficiency.

PVAC's victim-based counting enables larger \nbo configurations than PRAC while maintaining security under a lower victim-row \rh threshold (\nrhvic).
Here, \nrhvic denotes the maximum hammered count that a victim row can tolerate without inducing bitflips.
PVAC directly bounds the hammered count observed at each row, while PRAC must conservatively restrict each aggressor's hammering count to account for disturbance aggregation within the blast radius.
To quantify this difference, we model a worst-case feinting attack~\cite{sp-2022-protrr,aixiv-2016-silverbullet} and analytically derive the \nbo values for PVAC and PRAC to tolerate an identical maximum hammered count.

PVAC improves both performance and energy efficiency under benign workloads and malicious access patterns.
We evaluate its performance and energy consumption against PRAC and state-of-the-art PRAC-based mitigation schemes, including Chronus~\cite{hpca-2025-chronus}, MOAT~\cite{asplos-2025-moat}, and QPRAC~\cite{hpca-2025-qprac}.
Under benign workloads, PVAC consistently delivers higher performance than PRAC-based schemes while reducing energy consumption by avoiding unnecessary proactive mitigations and operating without long-latency PRAC timing parameters.
Even under the adversarial access patterns we evaluated, PVAC outperforms Chronus.

The key contributions of this paper are as follows: 
\begin{itemize}[leftmargin=*,nolistsep]
\item We identify fundamental drawbacks in aggressor-based counting and the PRAC organization, and we propose a victim-based counting method, Per-Victim-row hAmmered Counting (PVAC), to address these limitations.
\item PVAC enables larger \nbo than PRAC under an identical worst-case bound and remains safe at lower thresholds.
\item We show that PVAC incurs the lowest performance and energy overhead among PRAC-based schemes we evaluated. 
\end{itemize}

\section{Background}
\label{sec:2_background}

\subsection{DRAM operations and timing parameters}
\label{subsec:2_1_dram}

\begin{figure}[!tb]
  \center
  \includegraphics[width=0.96\columnwidth]{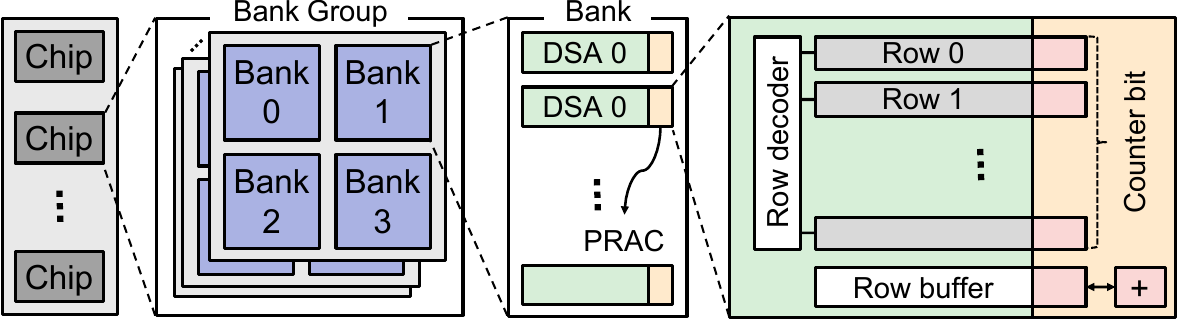}
  \vspace{-0.07in}
  \caption{A DRAM organization equipped with PRAC.}
  \label{fig:dram_org}
  \vspace{-0.09in}
\end{figure}

To access data in DRAM, a memory controller (MC) issues a sequence of DRAM commands with predefined timing constraints.
DRAM chips are organized hierarchically into bank groups, banks, and data subarrays (DSAs), which contain multiple rows and columns (Figure~\ref{fig:dram_org}).
To access data, the MC must first activate (\actcmd) the row containing the data, which enables a row decoder and loads the data into the row buffer. 
After that, read (\rdcmd) or write (\wrcmd) operations can be performed.
When accessing data from a different row, the currently activated row must be precharged (\precmd) before issuing a subsequent \actcmd command.
Table~\ref{tbl:dram_timings} summarizes the timing parameters that define the ordering and timing constraints among the commands.

\begin{table}[]
  \caption{Major timing parameters for DDR5-4800 without/with PRAC}
  \vspace{-0.05in}
  \label{tbl:dram_timings}
  \centering
  \begin{tabularx}{\columnwidth}{@{} p{0.8cm} p{5.1cm} r r @{}}
    \toprule
    \textbf{Timing} & \textbf{Description} & \textbf{Default} & \textbf{PRAC} \\
    \midrule
    \tras      & Min. time from \actcmd to \precmd            &  32\,ns   & 16\,ns  \\
    \trp       & Min. time from \precmd to \actcmd            &  16\,ns   & 36\,ns  \\
    \trc       & Min. time between consecutive \texttt{ACTs}            &  48\,ns   & 52\,ns  \\
    \trtp      & Min. time from \rdcmd to \precmd             & 7.5\,ns   &  5\,ns  \\
    \twr       & Min. time from the end of write burst to \precmd  &  30\,ns   & 10\,ns  \\
    \trcd      & Min. time from \actcmd to \rdcmd \
    / \wrcmd &  16\,ns & 16\,ns \\ 
    \bottomrule
  \end{tabularx}
  \vspace{-0.1in}
\end{table}

The timing parameters are largely determined by circuit-level properties.
In particular, \tras, \trp, \twr, and \trcd depend on the number of rows within a subarray.
Populating more rows increases the capacitance of the sensing circuitry, which in turn lengthens the sensing time~\cite{isca-2013-charm, tiered-2013-HPCA}.

To ensure data retention, an MC periodically issues refresh (\refcmd) commands, guaranteeing that each row is activated and precharged at least once within the refresh window (\trefw) to restore charge lost due to leakage.
Table~\ref{tbl:ref_timings} lists the timing parameters associated with \refcmd.
There are two types of refresh operations: \refabcmd refreshes all banks simultaneously, and \refsbcmd refreshes one bank per bank group. 
The MC issues an \refcmd command every refresh interval (\trefi), providing DRAM with a refresh period of \trfc, whose duration depends on the \refcmd type and DRAM chip capacity.

\subsection{RowHammer (\rh)}
\label{subsec:2_2_rowhammer}

\rh is a phenomenon in which repeatedly activating a DRAM row (an aggressor row) induces bitflips in its neighboring rows (victim rows).
We refer to the activation of an aggressor row as \emph{hammering}, and the victim rows that receive disturbance as \emph{hammered}. 
To distinguish the two perspectives, we use \nrhagg to denote the maximum hammering count allowed for each aggressor row, and \nrhvic to denote the maximum hammered count that a victim row can tolerate without inducing bitflips.
Because this threshold can vary depending on attack patterns and DRAM cell values~\cite{isca-2024-dramscope}, we define \nrhvic and \nrhagg using their minimum values.
Additionally, the disturbance can extend beyond the immediately adjacent rows to affect rows at greater distances---a behavior commonly referred to as the \emph{blast radius} (\br)---and JEDEC has taken the \br into account in its specifications~\cite{jedec-2024-ddr5}.

Although \nrhagg and \nrhvic are related, they capture fundamentally different quantities.
When multiple aggressor rows within the \br contribute to the same victim row, their effects accumulate at the victim.
Thus, aggressor-based mitigation schemes must conservatively set \nrhagg below \nrhvic so that the total hammered count at any victim row never exceeds \nrhvic.
As \br grows, the gap between \nrhagg and \nrhvic becomes larger.

To mitigate \rh, DRAM vendors have implemented in-DRAM Target Row Refresh (TRR) mechanisms starting since DDR4.
In-DRAM TRR implementations have adopted counter-based, sampling-based, or hybrid schemes that combine both approaches~\cite{micro-2021-uncovering,arxiv-2023-dsac,isscc-2023-hynixrowhammer}.
These mechanisms identify potential victim rows and use a portion of \trfc to refresh those victim rows.
While in-DRAM TRR reduces the impact of \rh, it remains vulnerable to carefully crafted adversarial access patterns~\cite{sp-2020-trrespass, sp-2022-blacksmith, micro-2021-uncovering}.

Moreover, the DDR5 JEDEC standard introduces Refresh Management (\rfmcmd) as a formal mitigation primitive~\cite{jedec-2024-ddr5}.
By issuing an \rfmcmd command, the MC grants DRAM a timing window during which the device may perform internal mitigative actions. 
Upon receiving the \rfmcmd command, DRAM can refresh potential victim rows using the allotted time.

\subsection{Per-Row Activation Counting (PRAC)}
\label{subsec:2_3_prac}

PRAC is a \rh mitigation mechanism that extends each DRAM row to include extra counter bits and performs exact activation counting by updating the counter value on every \precmd command~\cite{jedec-2024-ddr5}.
The activations include \actcmd, \refcmd, and \rfmcmd operations~\cite{jedec-2024-ddr5}. 
PRAC updates counter bits internally through a read-modify-write operation performed at every \precmd command.
This additional operation alters several timing parameters. 
Table~\ref{tbl:dram_timings} summarizes the timing parameters under both default and PRAC configurations. 
Among the timing parameters modified by PRAC, the longer \trp introduces performance overhead~\cite{isca-2025-mopac, hpca-2025-chronus, cal-2025-ohprac, arxiv-2025-practical}.

\begin{table}[tb!]
    \centering
    \caption{Refresh timing parameters for 16 {\normalfont Gb} DDR5 devices}
    \vspace{-0.05in}
    \label{tbl:ref_timings}
    \begin{threeparttable}
        \begin{tabularx}{0.99\columnwidth}{@{} l X r r r @{}}
            \toprule
            \textbf{Command} & \textbf{Refresh mode} & $\boldsymbol{t_{\mathrm{REFW}}}$ & $\boldsymbol{t_{\mathrm{REFI}}}$ & $\boldsymbol{t_{\mathrm{RFC}}}$ \\
            \toprule
            \refabcmd & Normal & 32 ms & 3.9\,$\mu$s & 295\,ns \\
            \refabcmd & Fine granularity & 32 ms & 1.95\,$\mu$s & 160\,ns \\
            \refsbcmd & Fine granularity & 32 ms & 1.95/4\,$\mu$s & 130\,ns \\
            \bottomrule
        \end{tabularx}
    \end{threeparttable}
    \vspace{-0.10in}
\end{table}

\begin{figure}[!tb]
  \center
  \includegraphics[width=0.93\columnwidth]{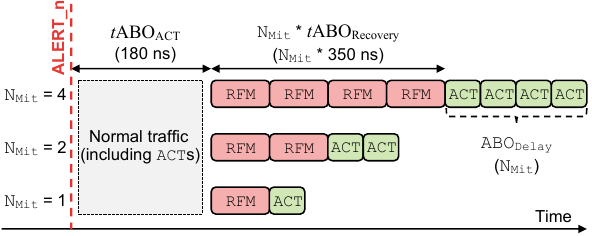}
  \vspace{-0.1in}
  \caption{Alert Back-Off (ABO) protocol overview.}
  \vspace{-0.1in}
  \label{fig:abo}
\end{figure}

When the activation count of a row exceeds the Back-Off threshold (\nbo), PRAC invokes the Alert Back-Off (ABO) protocol to mitigate potential \rh effects.
Figure~\ref{fig:abo} shows the overall sequence once DRAM asserts the ALERT\_n signal (\alert) to the MC.
Upon receiving the signal, the MC provides a \taboact timing margin before initiating mitigation.
During this period, up to \aboact activations can be issued.
After \taboact, the MC issues \nmit \rfmcmd commands, allowing DRAM to refresh potential victim rows internally.
Each \rfmcmd occupies 350\,ns, during which \actcmd commands are blocked, resulting in a total stall time of \taborec = \nmit $\times$ 350\,ns.
Within this duration, DRAM performs victim row refreshes and resets the counter value of the aggressor row.
After all \texttt{RFMs} are complete, the next \alert can be triggered only after \abodel activations have been issued.
Table~\ref{tbl:term} summarizes the PRAC terminologies.

To determine which rows should be refreshed, PRAC can maintain a queue structure that stores row addresses with their counter values~\cite{jedec-2024-ddr5}.
When any counter in this queue reaches a specific threshold, DRAM triggers an \alert.

Further, similar to the In-DRAM TRR in DDR4, PRAC can incorporate proactive mitigations that refresh potential victim rows in advance during \trfc~\cite{hpca-2025-qprac, hpca-2025-chronus, asplos-2025-moat}.
We classify \trfc operations into normal refresh, which periodically refreshes all rows to ensure data retention, and proactive mitigation, which resets an aggressor row counter and refreshes its victim rows, similar to an \rfmcmd operation.

\section{Threat Model}
\label{sec:3_threat_model}

We define a typical \rh threat model consistent with prior studies~\cite{hpca-2025-chronus,hpca-2025-qprac,asplos-2025-moat,isca-2025-mopac}.
We assume a \br of two, as in Chronus~\cite{hpca-2025-chronus}, which is also used in commodity DRAM devices~\cite{security-2025-eccfail}.
We assume that a powerful attacker can generate memory requests targeting specific DRAM components (\eg, banks, rows, and columns) and possesses full knowledge of the deployed defense algorithm.
Under this assumption, the attacker can craft adversarial memory access patterns for \rh attacks. 
Other disturbance attacks, such as RowPress~\cite{isca-2023-rowpress} and ColumnDisturb~\cite{micro-2025-coldist}, are out of scope.

\section{Motivation}
\label{sec:4_motivation}

\renewcommand{\arraystretch}{1.2}
\begin{table}  
  \caption{Terminologies related to PRAC}
  \vspace{-0.05in}
  \label{tbl:term}
  \centering
  \begin{tabularx}{\columnwidth}{@{} p{1.7cm} | p{5.1cm} | l @{}}
    \Xhline{2.5\arrayrulewidth}
    \textbf{Symbol} & \textbf{Description} & \textbf{Value} \\ 
    \Xhline{2.5\arrayrulewidth}
    \nrhagg / \nrhvic & Minimum aggressor hammering / victim hammered count for the first \rh bitflip &   \\
    \hline
    \nbo & Back-Off threshold &   \\
    \hline
    \nmit & The number of \texttt{RFM}s on \alert & 1, 2, or 4 \\
    \hline
    \taboact & Maximum time window from \alert to \rfmcmd & 180\,ns \\
    \hline
    \taborec & Time window for performing \texttt{RFM}s & 350\,ns \\
    \hline
    \aboact & The possible number of \texttt{ACT}s during \taboact & 3 \\
    \hline
    \abodel & The minimum number of \texttt{ACT}s allowed from end of \rfmcmd to the next \alert & \nmit  \\
    \hline
    \br & Blast radius & 2  \\
    \Xhline{2.5\arrayrulewidth}
  \end{tabularx}
  \vspace{-0.1in}
\end{table}
\renewcommand{\arraystretch}{1.0}

We investigate the necessity of \textit{victim-based counting} for mitigating \rh attacks.
\textit{Aggressor-based counting}, which increments a counter on every activation, overlooks the charge restoration naturally provided by activations, causing unnecessary counter accumulation.
Thus, PRAC suffers from performance degradation due to excessive \rfmcmd operations~\cite{hpca-2025-qprac,hpca-2025-chronus,asplos-2025-moat, isca-2025-mopac}.
Also, DRAM cell-based counters inherently differ from SRAM-based counters, imposing more constraints on their operations.
%

\subsection{Limitations of aggressor-based counting}
\label{subsec:4_1_assumptions}

Aggressor-based counting ignores the physical charge restoration that occurs when a row is activated.
While an activation physically restores the accessed row, PRAC blindly increments that row's hammering count (\S\ref{sec:1_introduction}).
This causes persistent, monotonic counter accumulation even for unaccessed rows, eventually triggering unnecessary \alerts.
Further, these \alerts exacerbate the problem: because \rfmcmd operations activate adjacent victim rows, mitigating one row increments its neighbors' counters, potentially triggering further \alerts in a cascading effect.

To prevent this accumulation, frequent counter resets are required, but they are prohibitively expensive in aggressor-based schemes.
Safely resetting an aggressor's counter requires first refreshing all neighboring victim rows.
Because PRAC stores counters within DRAM rows, this process requires at least five serialized activations (for a \br of two), incurring a severe latency penalty.
Consequently, PRAC is forced to delay resets until \alert-induced \rfmcmd operations, which provide a sufficiently long timing window (350\,ns).

\subsection{Structural Limits of DRAM Cell-based Counters}
\label{subsec:4_2_diff}

Unlike prior SRAM/CAM-based mitigations that permit parallel counter resets every \trefw~\cite{hpca-2022-mithril,isca-2019-twice,sp-2022-protrr,arxiv-2023-dsac,micro-2020-graphene}, PRAC embeds counters directly within DRAM rows.
This strict adherence to DRAM timing dictates that every counter update requires an explicit \actcmd and \precmd, making large-scale periodic resets infeasible.
Consequently, PRAC and its derivatives are forced to rely on expensive \alert-induced \rfmcmd or heavily extended normal refresh windows (\eg, MOAT~\cite{asplos-2025-moat} inflates \trfc from 295\,ns to 410\,ns) simply to accommodate the latency of counter resets.

\subsection{Need for victim-based counting}
\label{subsec:4_3_victim}

Victim-based counting resolves the reset bottlenecks of PRAC.
By resetting a row's counter upon its activation---which accurately reflects its physical charge restoration---and incrementing the adjacent victim rows, PVAC intrinsically prevents persistent accumulation and eliminates the multi-row structural overheads of aggressor-based tracking.

\renewcommand{\arraystretch}{1.4}
\begin{table}[t]
    \centering
        \caption{Comparison with prior PRAC studies}
        \label{tbl:prior}
        \begin{adjustbox}{width=\columnwidth,center}
        \begin{tabular}{l|ccccc}
        \Xhline{2.5\arrayrulewidth}
             & \makecell{PRAC\\\cite{jedec-2024-ddr5}} & \makecell{Chronus\\\cite{hpca-2025-chronus}} & \makecell{QPRAC\\\cite{hpca-2025-qprac}} & \makecell{MOAT\\\cite{asplos-2025-moat}} & \textbf{PVAC} \\
        \Xhline{2.5\arrayrulewidth}
            Target of counting & Aggr. & Aggr. & Aggr. & Aggr. & \textbf{Victim} \\
            Counter reset on \actcmd & No & No & No & No & \textbf{Yes} \\
            PRAC timing usage & Yes & No & Yes & Yes & \textbf{No} \\
        \Xhline{2.5\arrayrulewidth}
    \end{tabular}
    \end{adjustbox}
    \vspace{-0.05in}
\end{table}
\renewcommand{\arraystretch}{1.0}

\subsection{Prior works}
\label{subsec:3_3_prior_work}

Table~\ref{tbl:prior} categorizes prior PRAC-based studies~\cite{asplos-2025-moat,hpca-2025-qprac,hpca-2025-chronus} along three key aspects: counting target, counter reset behavior on an \actcmd operation, and PRAC timing usage.
Among these, all previous studies track activations of aggressor rows, which inherently prevents counter reset on \actcmd. 
Chronus~\cite{hpca-2025-chronus} is the only work that introduces a counter subarray that allows concurrent counter updates, enabling it to operate under the original DDR5 timing rather than PRAC timing.

In summary, none of the prior PRAC studies adopt a victim-based counting method that enables seamless counter resets during \actcmd operations.
Our goal is to implement victim-based counting PRAC without incurring additional overhead.

\section{Per-Victim-row hAmmered Counting}
\label{sec:5_pvac}

PVAC shifts the counting paradigm from aggressor rows to victim rows.
Every ACT inherently resets the accessed row's counter in PVAC, eliminating the need to activate multiple consecutive rows for a single reset and making refresh-coupled resets feasible.
However, because each \actcmd must modify multiple counters (\eg, the accessed row and four adjacent victims when \br$=2$), PVAC employs a dedicated counter subarray (CSA) to perform these updates concurrently without extending the critical timing path.
All evaluations use DDR5-4800 B-die 16Gb chips with 32 banks per sub-channel, 64K rows per bank, and 1\,KB row size (Table~\ref{tbl:dram_timings} and Table~\ref{tbl:ref_timings}).

\subsection{PVAC structure}
\label{subsec:5_1_pvac_structure}

\begin{figure}[!tb]
  \center
  \includegraphics[width=0.96\columnwidth]{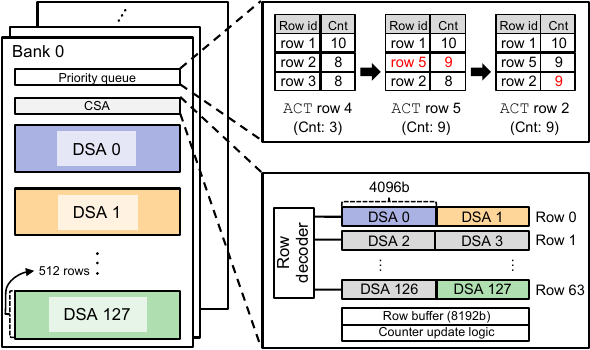}
  \vspace{-0.1in}
  \caption{An overview of PVAC architecture, including a counter subarray (CSA) and a priority queue.}
  \label{fig:pvac}
  \vspace{-0.1in}
\end{figure}

Figure~\ref{fig:pvac} illustrates the overall structure of PVAC, which consists of two main components: 1) a CSA and 2) a priority queue.
A DSA stores user data for normal memory access, whereas a CSA maintains the hammered counts of every row in each DSA.
We assume that each DSA contains 512 rows, and each row has an 8-bit counter, a commonly used configuration in prior works~\cite{hpca-2025-chronus, dramsec-2025-cncprac}.
The priority queue, as introduced in~\cite{hpca-2025-qprac}, is updated on every \actcmd and tracks the top-K victim rows with the highest hammered counts (where K denotes the queue depth).

\noindent
\textbf{Counter subarray:}
As established in \S\ref{subsec:4_2_diff}, PRAC's embedding of counters within DSA rows strictly ties updates to the slow DRAM timing parameters.
Because PVAC must update the counters of the accessed row and its adjacent victims on every \actcmd, a traditional PRAC layout would demand up to five serialized DSA activations (for \br$= 2$).
This adds approximately $4\times$\trc (\eg, 208\,ns) of latency per normal access, severely degrading performance.

\begin{figure}[!tb]
  \center
  \includegraphics[width=1\columnwidth]{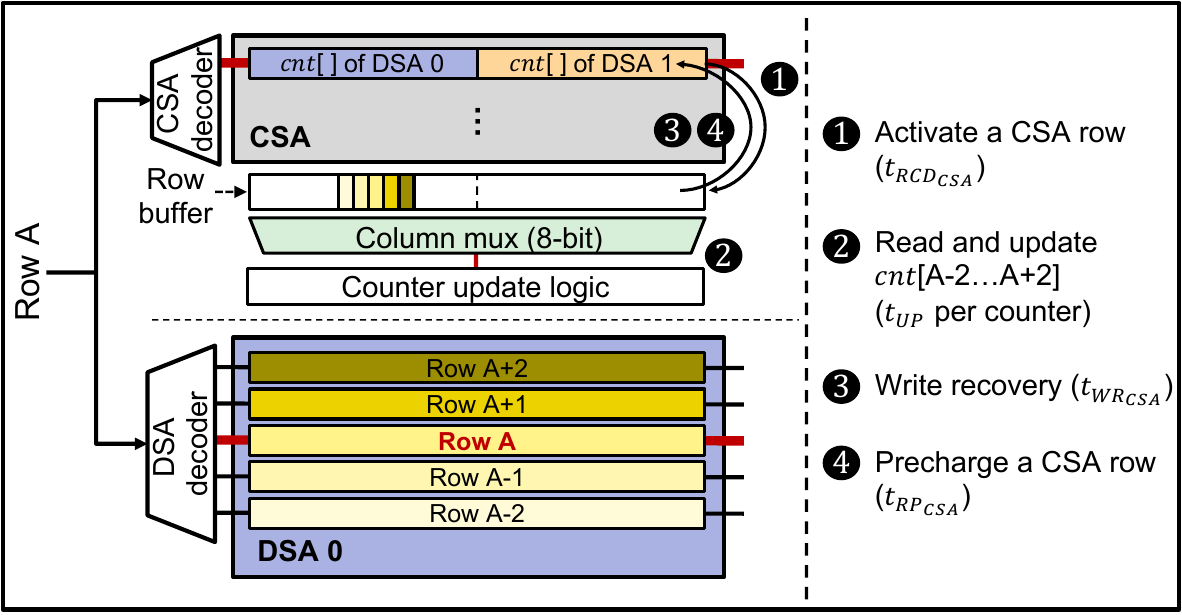}
  \vspace{-0.2in}
  \caption{Counter update sequence when a DSA row A is activated, with concurrent CSA row activation.}
  \vspace{-0.15in}
  \label{fig:update}
\end{figure}

To bypass these structural limits, PVAC extracts counter storage into a dedicated counter subarray (CSA), leveraging subarray-level parallelism~\cite{isca-2012-salp} as in prior studies~\cite{DRAMSec-2021-panopticon,hpca-2025-chronus}.
Because the CSA does not share the data path with the DSA, it can activate concurrently with DRAM accesses (Figure~\ref{fig:update}).
On a DSA \actcmd, the CSA simultaneously updates the adjacent victim counters and resets the accessed row's counter without extending the critical timing path.

Because \rh disturbance is confined within a single subarray~\cite{sosp-2023-siloz,msr-2021-mfit,isca-2024-dramscope,hotos-2021-stop,dsn-2024-hbm2-rowhammer}, PVAC maps the counters of each DSA to a single CSA row, enabling all counter updates with a single \actcmd.
Thus, PVAC need not examine or update counters in other DSAs.

The storage overhead of the CSA is negligible.
For 64K rows, each with an 8-bit counter, a CSA requires only 64\,KB (64 rows), \ie, less than 0.1\% capacity overhead~\cite{diephoto-2019-ISSCC}.
Given that a DSA typically contains 512 rows, each CSA row can store counter values for two DSAs.

Moreover, PVAC adopts guard rows~\cite{dimva-2018-guardion, osdi-2018-zebram, hpca-2022-safeguard, security-2017-catt} between CSA rows to prevent potential bitflips.
Assuming two guard rows per CSA row to account for a \br of two, the resulting capacity overhead is only 0.29\%~\cite{diephoto-2019-ISSCC}.
As PVAC performs counter resets during refresh operations, every CSA row is refreshed once per \trefw, thereby eliminating the need for additional refresh logic for the CSA.

\noindent
\textbf{Counter update logic:}
PVAC updates the counter values by simultaneously activating both a DSA row and a CSA row.
Figure~\ref{fig:update} illustrates the sequence of operations involved in counter updates.
Both the CSA and DSA decoders receive the input row address and activate the corresponding row.
Following prior work~\cite{hpca-2025-chronus}, we incorporate dedicated counter update logic within the CSA to update the counter efficiently.

Unlike PRAC, PVAC resets the counter of the activated row and updates the counters of its adjacent victim rows.
\trcdcsa, \twrcsa, and \trpcsa denote the CSA timing parameters corresponding to \trcd, \twr, and \trp of the DSA, respectively. 
\blackcircled{1}
When a DSA row is activated, the corresponding CSA row is activated concurrently.
\blackcircled{2}
After \trcdcsa, the counter update logic sequentially increments or decrements the counters of the four victim rows and then resets the counter of the activated row.
Each 8-bit counter value requires \tup to be read, updated, and written back; thus, a total latency of 5$\times$\tup is required.
\blackcircled{3} Once the update is finished, the CSA waits for \twrcsa.
\blackcircled{4} Finally, the row is precharged, which takes \trpcsa.

Each activation of a DSA row requires updating five counters—one for the activated row and four for its adjacent victim rows—but updating is completed within \trc because of the shorter CSA timing parameters.
As each CSA contains only 64 rows, its timing parameters
(\trcdcsa, \trascsa, \trpcsa, and \twrcsa)
are significantly smaller than the corresponding DSA timing parameters (\S\ref{subsec:2_1_dram})~\cite{isca-2013-charm, tiered-2013-HPCA, hpca-2017-soup}.
Based on the SPICE simulations from previous studies~\cite{isca-2020-clrdram,isca-2019-crow}, we apply reduction ratios (47.4\%, 52.0\%, 25.3\%, and 63.8\%) to the original DDR5 timing parameters, resulting in 7.6\,ns, 16.7\,ns, 4.1\,ns, and 19.2\,ns, respectively.
These values closely align with those reported in prior studies~\cite{isca-2013-charm,hpca-2017-soup}.

If the five counter values are updated serially, the total latency becomes \trcdcsa + 5$\times$\tup + \twrcsa + \trpcsa = (30.9 + 5$\times$\tup)\,ns, which fits within \trc of 48\,ns.
We synthesized the counter update logic using Synopsys Design Compiler~\cite{synopsys-design_compiler} with a TSMC 40nm process and measured \tup of 0.83\,ns.
Overall, the total counter update latency is 35.1\,ns and can be fully hidden within normal DSA operations.
The area overhead remains below 0.01\% per bank~\cite{diephoto-2019-ISSCC}, and the energy overhead is also negligible because CSA \actcmd and \precmd operations dominate the total update energy.

\noindent
\textbf{Priority queue:}
To determine which rows should be refreshed during proactive mitigation and \rfmcmd, consistent with previous work~\cite{hpca-2025-qprac,hpca-2025-chronus,isca-2025-mopac} and the JEDEC specification~\cite{jedec-2024-ddr5} (\S\ref{sec:2_background}), PVAC maintains a priority queue that stores row addresses together with their hammered counts, sorted by the count value.
%
Without such a queue, triggering an \alert would require scanning all activation counters to identify victim rows, leading to prohibitive performance overhead.
To avoid this overhead, PVAC maintains a per-bank priority queue so that mitigation candidates are readily available without requiring full-bank counter scans at each \alert.

Whenever a row is activated, PVAC simultaneously updates its counter value and determines whether the corresponding entry in the queue should be inserted or updated.
If the row already exists in the queue, only its hammered count is updated.
If a row's hammered count exceeds that of an existing entry, the entry with the smallest count is evicted and replaced with the new row.
By doing so, the queue always contains the top-K rows with the highest hammered counts.

\subsection{Mitigative actions}
\label{subsec:5_2_mitigate}
PVAC employs two commonly adopted mitigative actions to guarantee timely protection of victim rows: proactive mitigation and \rfmcmd.
Proactive mitigation is performed within \trfc to refresh potential victim rows, along with the normal refresh.
\rfmcmd is triggered when the hammered count of a specific victim row exceeds \nbo, generating an \alert.

\noindent
\textbf{Proactive mitigation:}
PVAC's proactive mitigation reduces the frequency of \alert invocations by refreshing the most hammered victim rows before they reach \nbo.
Similar to prior studies~\cite{micro-2021-uncovering,arxiv-2023-dsac}, PVAC leverages a portion of \trfc to selectively refresh the rows with the highest counter values in the priority queue.
To further enhance energy efficiency, PVAC does not invoke this mitigative action every \trfc; instead, it is triggered only when at least one row's counter value exceeds a predefined threshold (\eg, {\large $\sfrac{\nbo}{2}$})~\cite{hpca-2025-qprac}.
This selective triggering reduces the number of \texttt{Alerts} proactively.
Consistent with prior aggressor-based schemes that refresh four victim rows and reset one aggressor row per mitigation, PVAC performs mitigation by refreshing four victim rows.

\noindent
\textbf{RFM:}
Because PVAC increments up to four victim counters per ACT, multiple rows may exceed \nbo simultaneously.
To ensure all \alerts from a single activation are safely resolved, PVAC leverages the standard \rfmcmd timing window---which inherently accommodates five row accesses (four victims and one aggressor)---to refresh the four highest-priority victim rows directly from the priority queue.

\noindent
\textbf{Priority queue size:}
PVAC provisions its priority queue size to accommodate all the rows that may be refreshed during both \rfmcmd and proactive mitigation, ensuring correctness under the worst-case activation patterns.
The number of rows refreshed during \rfmcmd is the product of the maximum number of \texttt{RFMs} per \alert (\nmit), which can be up to four (Table~\ref{tbl:term}), and the number of rows refreshed per RFM (\ie, four).
Proactive mitigation adds up to four additional rows, totaling a queue size of 20.
This is four times larger than QPRAC’s queue size, but still incurs only 0.2\% area overhead per chip~\cite{hpca-2025-qprac}.

\subsection{Overhead of PVAC on normal refresh}
\label{subsec:5_3_energy_overhead}

While PVAC can perform counter resets during normal refresh operations, each refresh triggers additional counter updates and thus increases CSA energy consumption.
When rows in multiple DSAs are refreshed concurrently, their corresponding CSA rows must also be activated to update the associated counters.
For example, because each \refabcmd command refreshes eight rows per \trefi, refreshing a single row in eight even-indexed DSAs (\eg, DSA~0, 2, 4, ..., 14) requires activating up to eight CSA rows (Figure~\ref{fig:pvac}).

Further, because these CSA rows reside in the same subarray, they must be activated serially, resulting in non-trivial latency overhead in addition to the increased energy cost.
For example, activating eight CSA rows incurs a latency of 280.8\,ns $=(8 \times 35.1$\,ns), which occupies a critical portion of \trfc (295\,ns).
As proactive mitigation must also be completed within the \trfc, such long latency poses an additional critical challenge.
To make PVAC practical, it is essential to reduce the number of CSA activations for normal refreshes.

\subsection{CSA optimization for normal refresh}
\label{subsec:5_4_e-pvac}

\begin{figure}[!tb]
  \center
  \includegraphics[width=0.96\columnwidth]{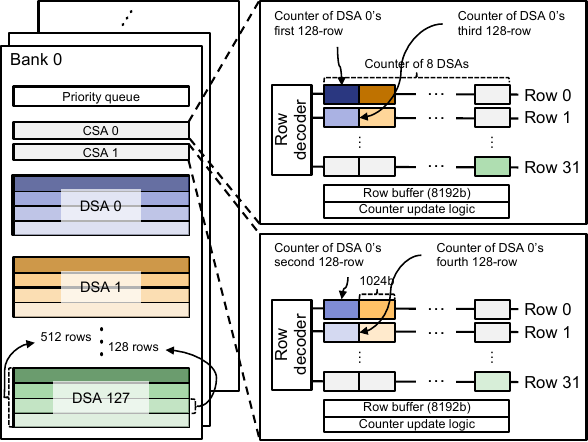}
  \vspace{-0.1in}
  \caption{Energy efficient CSA structure in PVAC, interleaving counters across dual CSAs to reduce overhead during normal refresh.}
  \vspace{-0.2in}
  \label{fig:e-pvac}
\end{figure}

To minimize energy consumption from CSA during normal refresh, we optimize a CSA structure to activate fewer CSA rows while preserving all functional behaviors of PVAC.
Baseline PVAC stores all counters of a DSA within a single CSA row, which requires activating many CSA rows when multiple DSAs are refreshed in parallel.
Instead, we map multiple DSAs' counter values into a single CSA row, while distributing each DSA’s counters across several CSA rows.
This remapping allows a single CSA row activation to update the counters of many DSA rows refreshed in parallel, substantially reducing CSA activation energy and latency.

However, distributing each DSA's counters across multiple CSA rows introduces a corner case.
When the required counters for a DSA row are spread over two CSA rows, the update requires two CSA activations, causing the operation to exceed \trc
(\eg, when row 128 is activated, the counter values of rows 126 through 130 must be updated).
Updating these five counters requires activating two CSA rows, and the resulting serialized activations incur 70.2\,ns $=(2 \times 35.1$\,ns) $>$ \trc.

To resolve this, we partition the original 64-row CSA into two parallel 32-row subarrays and interleave the counter storage.
The counters of each DSA are divided into four chunks, each containing the counters of 128 rows, and these chunks are interleaved across two CSAs (Figure~\ref{fig:e-pvac}).
That is, the counters for rows 126 and 127 are mapped to a different CSA than those for rows 128, 129, and 130. 
Hence, the activations are distributed across the two CSAs and can proceed in parallel, ensuring that all counter updates complete within \trc.

As dual-CSA activation occurs only in a small fraction of cases ($\sfrac{3}{128}$), this design reduces the additional energy cost of counter resets in refresh operations (\S\ref{subsec:7_3_perf}).
However, splitting the CSA introduces additional peripheral circuitry.
Since a sense amplifier is typically about two orders of magnitude larger than a DRAM cell~\cite{rambus-2016-drampower}, this additional peripheral logic inevitably incurs area overhead.
Nevertheless, the chip-level impact remains negligible, making the structure a practical and efficient enhancement to PVAC.

\section{Security Analysis}
\label{sec:6_security_analysis}

For a fair comparison between aggressor- and victim-based schemes, we evaluate \nbo values of all schemes under a common victim-side maximum hammered count (HC) constraint.
For aggressor-counting schemes, the \nbo values reported in this paper are not taken from their native \nrhagg-based analysis~\cite{hpca-2025-qprac,hpca-2025-chronus}.
Instead, we reformulate the analysis from the victim-row perspective and re-derive their \nbo values under the same maximum HC constraint.
Our analysis builds on the methodology of \cite{hpca-2025-qprac}, but focuses on victim rows' \emph{hammered count} rather than aggressor rows' \emph{hammering count}.
The \nbo value must be chosen such that the maximum HC of any victim row remains below \nrhvic.

We assume that the system employs the ABO protocol, which uses the \rfmcmd command as the only reactive mitigation mechanism.
Although proactive mitigation (and normal refresh in the case of PVAC) is also applied as a mitigative action, these mechanisms involve refresh postponement effects.\footnote{With PRAC enabled, refresh scheduling flexibility is reduced, limiting the maximum postponement from 5$\times$\trefi to 3$\times$\trefi~\cite{jedec-2024-ddr5}.}
For analytical clarity, we exclude such mechanisms from our model---representing a conservative assumption that places the defender in the worst-case condition.
Further, we assume that the attacker performs the \rh attack within a single \trefw, during which every DRAM row is refreshed exactly once.

\subsection{Feinting attack~\cite{sp-2022-protrr}}
\label{subsec:6_1_feinting}
To derive the worst-case HC achievable against PVAC and PRAC-based schemes, we adopt the feinting attack model~\cite{sp-2022-protrr} (or wave attack~\cite{aixiv-2016-silverbullet}), which maximizes HC within a single \trefw.
This attack proceeds in multiple rounds.
In each round, the attacker evenly distributes activations across a prepared set of rows.

When a counter exceeds \nbo, an \alert triggers \rfmcmd.
Assuming the attacker knows which rows are mitigated (\S\ref{sec:3_threat_model}), they exclude these from subsequent rounds and continue evenly activating the remaining rows until only one survives, experiencing the maximum HC.

\subsection{Attack modeling for PVAC}
\label{subsec:6_2_pvac_modeling}

\begin{figure}[!tb]
  \center
  \includegraphics[width=0.91\columnwidth]{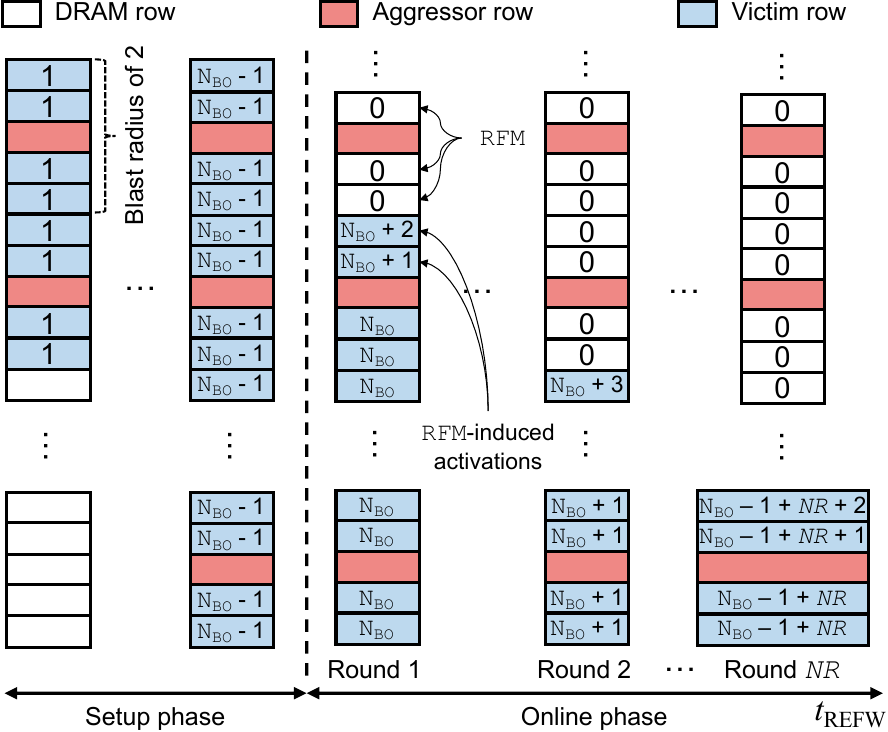}
  \caption{A sequence of feinting attack under a victim-based tracking mechanism with \br of 2~\cite{sp-2022-protrr}.}
  \label{fig:feinting}
  \vspace{-0.2cm}
\end{figure}

Figure~\ref{fig:feinting} illustrates the feinting attack on a victim-based counting scheme~\cite{sp-2022-protrr}. 
We analyze the feinting attack against PVAC by decomposing it into two phases: (1) \textit{setup phase} and (2) \textit{online phase}.
The total HC is obtained as the sum of HC from both phases:
\begin{equation} 
\vspace{-0.2em} 
\label{eq:totalhc} 
\Scale[0.85]{ 
\begin{split} 
\mathbf{HC} &= \mathbf{HC}_{setup} + \mathbf{HC}_{online}
\end{split}} 
\vspace{-0.2em} 
\end{equation}

\noindent
\textbf{Setup phase:}
During the setup phase, the attacker prepares an initial victim row pool ($R_{1}$) and issues $\nbo - 1$ activations to each victim's corresponding aggressor rows.
To efficiently increase the row pool size, the attacker activates rows with a stride of 5, considering a \br of 2~\cite{sp-2022-protrr}.
Therefore, the HC achieved in the setup phase is expressed as:
\begin{equation} 
\vspace{-0.2em} 
\label{eq:hcsetup_pvac} 
\Scale[0.85]{ 
\begin{split} 
\mathbf{HC}_{setup} &= \mathbf{N}_{BO} - 1
\end{split}} 
\vspace{-0.2em} 
\end{equation}

\noindent
\textbf{Online phase:}
The online phase proceeds over $NR$ rounds, where $NR$ denotes the number of attack rounds.
In each round, the attacker activates each remaining row once.
Rows within the \br of the refreshed rows experience additional \rfmcmd-induced activations (Figure~\ref{fig:feinting}).
In the final round ($NR$), the attacker can perform \aboact + \abodel activations before the last \rfmcmd commands are issued.
Therefore, the online phase HC can be expressed as:
\begin{equation} 
\vspace{-0.2em} 
\label{eq:hconlie_pvac} 
\Scale[0.85]{ 
\begin{split} 
\mathbf{HC}_{online} &= \mathbf{NR} + \mathbf{ABO}_{Delay} + \mathbf{ABO}_{ACT} + \mathbf{BR}
\end{split}} 
\vspace{-0.2em} 
\end{equation}
\noindent
Because all values except $NR$ are constant, the HC reaches its maximum when $NR$ reaches its highest possible value.

\noindent
\textbf{Determining NR:}
In round $n$, all remaining rows ($R_{n}$) are evenly hammered.
Based on the ABO protocol constraints (Table~\ref{tbl:term}), each \alert permits exactly \aboact + \abodel activations before the defense enforces the \rfmcmd penalty,
implying that exactly \aboact + \abodel rows in the pool can be activated once per \alert.

During a round, \texttt{Alerts} are triggered, and the resulting \rfmcmds refresh a subset of rows.
The number of \texttt{Alerts} generated in round $n-1$ is $\lfloor \frac{R_{n-1}}{4\times(ABO_{ACT} + ABO_{Delay})} \rfloor$.
Multiplying this by the number of victim rows refreshed per \rfmcmd (\ie, 4) and by \nmit yields the total number of mitigated rows.
Thus, the remaining row pool size for the next round ($R_{n}$) can be expressed as: 
\begin{equation}
\vspace{-0.2em}
\label{eq:rowpool_pvac}
\Scale[0.85]{
\begin{split}
\mathbf{R}_{n} &= \mathbf{R}_{n-1} - \mathbf{N}_{Mit} \times\lfloor\frac{\mathbf{R}_{n-1}}{\mathbf{ABO}_{ACT} + \mathbf{ABO}_{Delay}}\rfloor
\end{split}}
\vspace{-0.1em}
\end{equation}
\noindent

\noindent
$NR$ is the value of $n$ where $R_{n}=1$.

\subsection{Attack modeling for PRAC~\cite{hpca-2025-qprac}}
\label{subsec:6_3_prac_modeling}

As PRAC employs aggressor-based counting, its \nbo must be conservatively determined by considering \br.
Consider five contiguous rows (row 0--4).
If rows 0, 1, 3, and 4 are each activated $\sfrac{N}{4}$ times, their individual \emph{hammering counts} remain $\sfrac{N}{4}$, but the cumulative hammered count at row 2 reaches $N$.
PVAC can tolerate this behavior by ensuring that the hammered count does not exceed $N$.
In contrast, PRAC enforces a stricter bound that limits each aggressor's hammering count to at most $\sfrac{N}{4}$.
Even if only one of these rows is activated $\sfrac{N}{4}$ times, an \alert must be issued.
This gap in the threshold values between PRAC and PVAC increases as \br grows.
To reconcile this asymmetry, we derive the \nbo value for PRAC that guarantee the same maximum HC as PVAC.

We model the feinting attack on PRAC following \cite{hpca-2025-qprac} with one key difference in the layout of aggressor and victim rows.
To maximize the HC of a victim row, the aggressor rows must be placed within the victim row's \br so that their disturbance accumulates onto the victim.
With \br of 2, this requires the final surviving aggressor rows to surround one victim row, with two aggressors on each side.
As the victim is refreshed when any of those aggressors is mitigated, the attack ends when the remaining aggressor row pool $R_{n}$ becomes four.
Hence, $NR$ corresponds to the round index $n$ at which $R_n = 4$.

\noindent
\textbf{Setup phase:}
During the setup phase, each of the final $2\times$\br aggressor rows is prepared up to \nbo$-1$.
Because all of them lie within the same victim row's \br, their disturbance accumulates at the victim, yielding:
\begin{equation} 
\vspace{-0.3em} 
\label{eq:hcsetup_prac} 
\Scale[0.85]{ 
\begin{split} 
\mathbf{HC}_{setup} &= 2\times\mathbf{BR}\times(\mathbf{N}_{BO} - 1)
\end{split}} 
\vspace{-0.3em} 
\end{equation}

\noindent
\textbf{Online phase:}
In every round, the progression is the same as that of PVAC, except that the per-round HC contribution becomes $2\times$\br.
After the last round ($NR$), the attacker can further issue $ABO_{Delay}+ABO_{ACT}$ activations before \rfmcmd begins.
Then, as rows are refreshed in the order of distance 1, 2, \ldots, \br from the mitigated aggressor, the farthest victim row is refreshed last and can receive up to \br-1 additional disturbances before being refreshed.
Therefore, the online-phase HC is expressed as: 
\begin{equation} 
\vspace{-0.2em} 
\label{eq:hconlie_prac} 
\Scale[0.77]{ 
\begin{split} 
\mathbf{HC}_{online} &= (2\times\mathbf{BR}\times\mathbf{NR}) + \mathbf{ABO}_{Delay} + \mathbf{ABO}_{ACT} + \mathbf{BR} - 1
\end{split}} 
\vspace{-0.2em} 
\end{equation}

\noindent
\textbf{Determining NR:}
As in prior work~\cite{hpca-2025-qprac}, aggressor rows in PRAC are placed contiguously, since PRAC tracks only aggressor hammering counts and thus does not require the stride used in PVAC.
Because each \rfmcmd refreshes victim rows within \br of the mitigated aggressor, the last \rfmcmd in round $n-1$ effectively pre-activates the first \br rows of the aggressor row pool in round $n$~\cite{hpca-2025-qprac}.
Therefore, only $R_{n-1}-\br$ rows need to be considered when counting the number of \alert events in round $n-1$, yielding:
\begin{equation}
\vspace{-0.2em}
\label{eq:rowpool_prac}
\Scale[0.85]{
\begin{split}
\mathbf{R}_{n} &= \mathbf{R}_{n-1} - \mathbf{N}_{Mit} \times\lfloor\frac{(\mathbf{R}_{n-1} - \mathbf{BR})}{\mathbf{ABO}_{ACT} + \mathbf{ABO}_{Delay}}\rfloor
\end{split}}
\vspace{-0.2em}
\end{equation}
\noindent
For \br = 2, $NR$ is the round index at which only the final four aggressor rows around the victim remain (\ie, $R_n=4$).

\subsection{Attack modeling for Chronus}
\label{subsec:6_4_chronus_modeling}

Chronus proposes an optimized ABO protocol that enables a single \alert to mitigate all rows whose counter values exceed \nbo, regardless of \nmit~\cite{hpca-2025-chronus}. 
Accordingly, after the setup phase, one additional activation is needed to trigger the \alert, after which the attacker can issue \aboact more activations before \rfmcmd begins.
The farthest victim row can then receive up to $\br-1$ additional disturbances during \rfmcmd before being refreshed.
Therefore, the online-phase HC is $1+\aboact+(\br-1)=\aboact+\br$.
Since the setup phase is identical to that in Equation~\ref{eq:hcsetup_prac} and the online phase does not proceed over multiple rounds, we obtain:
\begin{equation} 
\vspace{-0.2em} 
\label{eq:hconlie_chronus} 
\Scale[0.85]{ 
\begin{split} 
\mathbf{HC}_{online} &= \mathbf{ABO}_{ACT} + \mathbf{BR}
\end{split}} 
\vspace{-0.2em} 
\end{equation} 
\noindent
and the total HC is:
\begin{equation}
\vspace{-0.2em}
\label{eq:hctotal_chronus}
\Scale[0.85]{
\begin{split}
\mathbf{HC} &= 2\times\mathbf{BR}\times(\mathbf{NBO}-1) + \mathbf{ABO}_{ACT} + \mathbf{BR}.
\end{split}}
\vspace{-0.5em}
\end{equation}

\begin{figure}[!tb]
  \center
  \includegraphics[width=0.94\columnwidth]{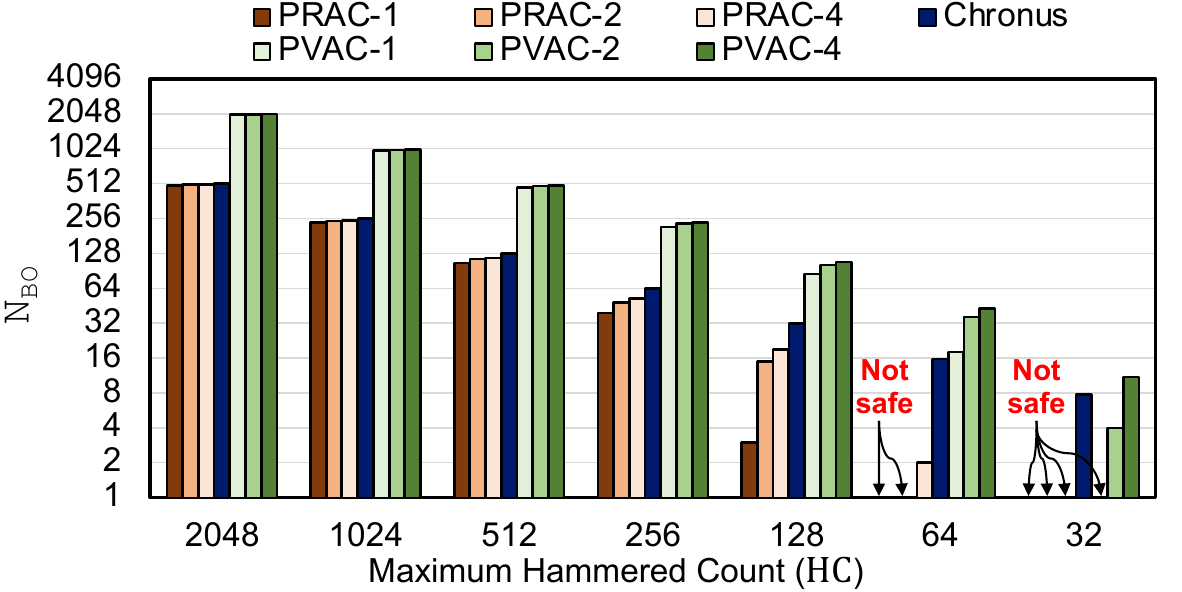}
  \vspace{-0.1in}
  \caption{\nbo across various maximum hammered count (HC) for PRAC, PVAC, and Chronus.}
  \vspace{-0.1in}
  \label{fig:security}
\end{figure}

\subsection{Result}
\label{subsec:6_4_result}

Figure~\ref{fig:security} illustrates \nbo as a function of maximum HC, when a bank consists of 64K rows, corresponding to a DDR5 16\,Gb configuration, and \nmit is set to 1, 2, or 4.
While the minimum and maximum initial victim row pool size under PVAC are 4 and 52,428, respectively, due to its stride-5 layout, the minimum and maximum initial aggressor row pool size of PRAC are 1 and 65,535, as one row must be reserved as the final victim row.
To compute $NR$, we solve Equation~\ref{eq:rowpool_pvac} and Equation~\ref{eq:rowpool_prac} for each initial row pool size ($R_{1}$) from minimum to maximum value.
We then substitute the resulting $NR$ into Equation~\ref{eq:hconlie_pvac} and Equation~\ref{eq:hconlie_prac} to obtain the corresponding HC values, and finally determine the \nbo values for PVAC and PRAC-based schemes under the same maximum HC.
For Chronus, only Equation~\ref{eq:hcsetup_prac} and Equation~\ref{eq:hconlie_chronus} need to be solved.

PVAC can be configured with larger \nbo values than PRAC-based schemes.
When the maximum HC is 128 and \nmit is set to 1, 2, and 4, the resulting \nbo values for PVAC are 85, 102, and 108, respectively, whereas those for PRAC are only 3, 15, and 19, and Chronus exhibits 31.
At the maximum HC of 2048, PVAC exhibits \nbo values of 2015, 2028, and 2032 for \nmit = 1, 2, and 4, respectively, whereas PRAC achieves 495, 501, and 503.
For Chronus, the derived \nbo value is 511.

PVAC can guarantee safety at lower maximum HC values than PRAC.
At a maximum HC of 64, PRAC-1 and PRAC-2 fail to accommodate this value, and only PRAC-4 can be configured with \nbo of 2.
Because PRAC must conservatively configure \nbo, it requires a much smaller value.
Consequently, as the maximum HC decreases, \nbo approaches zero earlier than PVAC, leading to unsafe points.
In contrast, PVAC remains safe with \nbo set to 43 at a maximum HC of 64 and continues to do so even when the maximum HC is reduced to 32, where \nbo is 11.
Chronus can also tolerate these points with \nbo of 15 and 7, respectively, due to its optimized ABO protocol (\S\ref{subsec:6_4_chronus_modeling}).

\section{Evaluation}
\label{sec:7_evaluation}
In this section, we evaluate the performance and energy overhead of PVAC in comparison with PRAC and other PRAC-based mitigation schemes.
We sweep the maximum HC values and, for each HC setting, compare all mitigation schemes under the same maximum HC constraint.

\subsection{System configuration}
\label{subsec:7_1_config}

{
\renewcommand{\arraystretch}{1.4}
\setlength{\tabcolsep}{8pt}
\setlength{\arrayrulewidth}{0.5pt}  
\begin{table}[t]
    \centering
        \caption{Configurations of mitigations}
        \vspace{-0.1in}
        \label{tbl:config}
        \begin{adjustbox}{width=\columnwidth,center}
        \Large
        \begin{tabular}{l|cccccc}
        \Xhline{2\arrayrulewidth}
         \textbf{Schemes} & \textbf{PRAC} & \makecell{\textbf{Chronus} \\ \cite{hpca-2025-chronus}} & \makecell{\textbf{QPRAC} \\ \cite{hpca-2025-qprac}} & \makecell{\textbf{MOAT} \\ \cite{asplos-2025-moat}} & \textbf{PVAC} \\
        \Xhline{2\arrayrulewidth}
        Proactive threshold & - & - & \nbo~/ 2 & \nbo~/ 2 & \nbo~/ 2 \\
        Proactive period & - & 2$\times$\trefi & 1$\times$\trefi & 4$\times$\trefi & 1$\times$\trefi \\
        \trfc (ns) & 295 & 295 & 295 & 410 & 295 \\
        \nmit & 1/2/4 & Adaptive & 1/2/4 & 1 & 1/2/4 \\
        \Xhline{2\arrayrulewidth}
    \end{tabular}
    \end{adjustbox}
    \vspace{-0.1in}
\end{table}
}

Table~\ref{tbl:config} summarizes the configurations of PVAC and the compared PRAC-based schemes.
For aggressor-counting schemes, the \nbo values used here are not directly taken from their native \nrhagg-based configurations, but are re-derived under the victim-side HC formulation described in Section~\ref{sec:6_security_analysis}.
The proactive threshold and period denote the counter value threshold and period that trigger proactive mitigations, respectively.
Chronus does not employ a proactive threshold.
Instead, it always performs proactive mitigation once every two \trefi intervals.
QPRAC, MOAT, and PVAC all adopt a threshold of {\large $\sfrac{\nbo}{2}$}.
Among them, QPRAC and PVAC execute proactive mitigation every \trefi, whereas MOAT performs it once every four \trefi by default.

All schemes refresh four victim rows per proactive mitigation.
PVAC resets the counter values of those victim rows, while the others reset the counters of a corresponding aggressor row.
Consistent with prior studies, conventional PRAC is assumed not to perform proactive mitigation~\cite{hpca-2025-qprac, hpca-2025-chronus}.
MOAT assumes a \trfc of 410\,ns, while the others use 295\,ns.

Chronus adaptively issues \rfmcmd commands for all aggressor rows whose counter values exceed \nbo within a single \alert. 
As MOAT is designed for the case of \nmit = 1, we compare it under that configuration~\cite{asplos-2025-moat}.
The other schemes support \nmit values of 1, 2, or 4. 
For each HC value, we evaluate all feasible \nmit configurations for each respective scheme and report the configuration that satisfies the HC constraint with the lowest overhead.

\begin{table}[t]
    
    \centering
        \caption{System configuration}
        \vspace{-0.1in}
        \label{tbl:system}
        \begin{adjustbox}{width=\columnwidth,center}
        \begin{tabular}{l|l}
        \toprule
        Processor & \makecell{4.2\,GHz, 4-core, 128-entry instruction window} \\
        \midrule
        Last-level cache & \makecell{64\,B cache line, 8-way set associative, 8\,MB} \\
        \midrule
        Memory controller & \makecell{64-entry read/write queue, FR-FCFS~\cite{isca-2000-fr-fcfs}, \\ MOP4CLXOR address mapping~\cite{micro-2011-minimalist}} \\
        \midrule
        Main memory & \makecell{DDR5-4800, 2 sub-channels, 1 rank, 8 bank groups, \\ 4 banks/bank group, 64K rows/bank, 1KB/chip} \\
        \bottomrule
    \end{tabular}
    \end{adjustbox}
    \vspace{-0.15in}
\end{table}

\subsection{Methodology}
\label{subsec:7_2_meth}

\noindent
\textbf{Simulation.}
We evaluate performance and energy consumption using Ramulator 2.0~\cite{cal-2024-ramulator2}, a cycle-accurate memory system simulator.
Table~\ref{tbl:system} summarizes the system configuration used in our experiments.
We assume a four-core out-of-order system with an 8\,MB last-level cache (LLC).
The main memory consists of two sub-channels, each with 32 banks, totaling 64 banks of DDR5-4800.
Accordingly, the LLC capacity and the number of banks per core are 2\,MB and 16, respectively.
The MC employs FR-FCFS scheduling policy~\cite{isca-2000-fr-fcfs}.

\noindent
\textbf{Workloads.}
We evaluated PVAC and prior work using five benchmark suites: SPEC CPU2006~\cite{spec-cpu-2006}, SPEC CPU2017~\cite{spec-cpu-2017}, TPC~\cite{tpc-benchmark}, MediaBench~\cite{Mediabench-benchmark}, and YCSB~\cite{ycsb-benchmark}.
We classify these workloads into three groups---High, Mid, and Low---based on their Row-Buffer Misses Per Kilo Instructions (RBMPKI)~\cite{cal-2025-ohprac,hpca-2021-blockhammer,hpca-2024-comet,hpca-2025-chronus}.
Using these classifications, we construct 30 mixed workloads, with 10 from each group. 
%

\subsection{Benign workloads}
\label{subsec:7_3_perf}

\begin{figure}[!tb]
  \center
  \includegraphics[width=0.96\columnwidth]{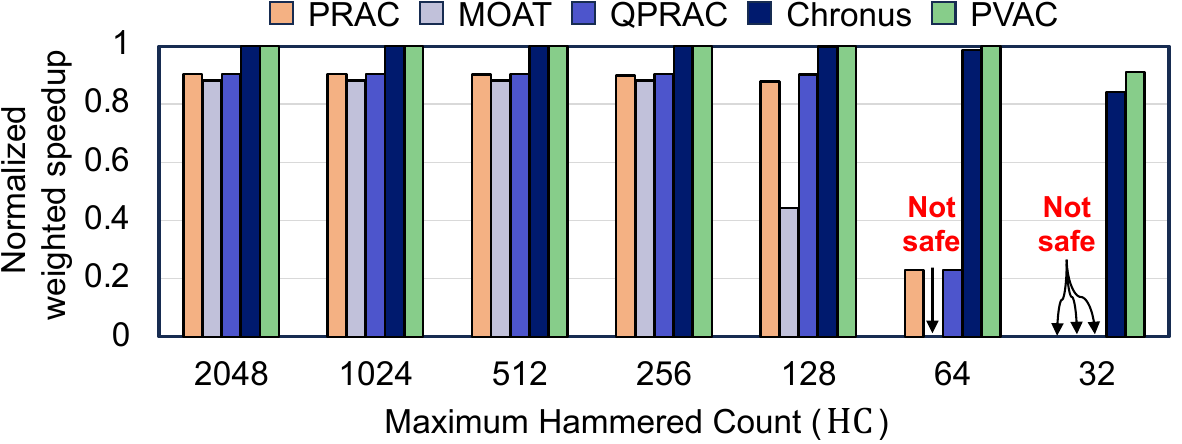}
  \vspace{-0.07in}
  \caption{Normalized weighted speedup across different workloads and HC values.}
  \label{fig:benign_perf}
  \vspace{-0.15in}
\end{figure}

\noindent
\textbf{Performance.}
Figure~\ref{fig:benign_perf} shows the performance impact under different HC constraints. 
For each group, we compute weighted speedup, normalize it to the baseline with no mitigation, and report the geometric mean across all mixed workloads.
We sweep HC from 32 to 2048 and evaluate both the na\"ive and optimized versions of CSA.
Because the CSA optimization does not impact performance, we report a single representative result.

When HC is below 32, the required \nbo values become extremely small to remain secure. 
For example, Chronus must set \nbo to 3 at HC = 16, and to 1 at HC = 8, to satisfy the security constraint.
Such configurations are no longer practically operable under benign workloads.
For this reason, we present evaluation results for HC values of 32 and above.

PVAC and Chronus achieve the highest performance among all the evaluated schemes.
Both schemes utilize dedicated CSAs, thereby operating with the default timing parameters (Table~\ref{tbl:dram_timings}).
In contrast, the other schemes adopt PRAC's timing parameters, which incur additional memory access latency and degrade performance~\cite{hpca-2025-chronus,cal-2025-ohprac, isca-2025-mopac}.

The performance gap between PVAC/Chronus and the other mitigation schemes further widens as HC decreases. 
PVAC employs a larger \nbo and resets counters on every access.
Chronus benefits from its adaptive RFM behavior that limits mitigation frequency.
These properties enable PVAC and Chronus to sustain higher performance while remaining secure even at lower HC values.
At lower HC values, PVAC outperforms Chronus by generating fewer \alerts under all benign workloads.
For the same HC value, PVAC, which counts victims, can configure a higher \nbo than Chronus, which counts aggressors, resulting in a larger performance gap at lower HC settings.
Specifically, when HC is 64, PVAC achieves a 1.3\% performance improvement over Chronus, and when HC is 32, the performance gap widens to 6.9\%.

Among the mitigations adopting PRAC timing parameters, performance decreases in the order of QPRAC, PRAC, and MOAT. 
PRAC issues more \rfmcmd commands than any other schemes, and it triggers \alerts even in the high group when HC is 2048, exhibiting the most significant \rfmcmd-induced performance reduction.
It is because PRAC neither resets counters during normal refresh nor employs proactive mitigation.

QPRAC outperforms PRAC due to its proactive mitigation.
At an HC of 256, QPRAC does not trigger an \alert, while PRAC incurs \alert-induced performance degradation. 
QPRAC triggers \alert only when HC falls below 128. 

MOAT generally performs worst. 
Similar to QPRAC, MOAT employs proactive mitigation; thus, it rarely triggers \texttt{Alert}s until HC reaches 256.
However, its \trfc (410\,ns) is longer than that of other schemes (295\,ns), resulting in higher refresh overhead.
Further, as MOAT only supports \nmit = 1, it must configure a lower \nbo than QPRAC and PRAC, which support \nmit up to 4, under the same HC constraint (Figure~\ref{fig:security}).

\begin{figure}[!tb]
  \center
  \includegraphics[width=0.98\columnwidth]{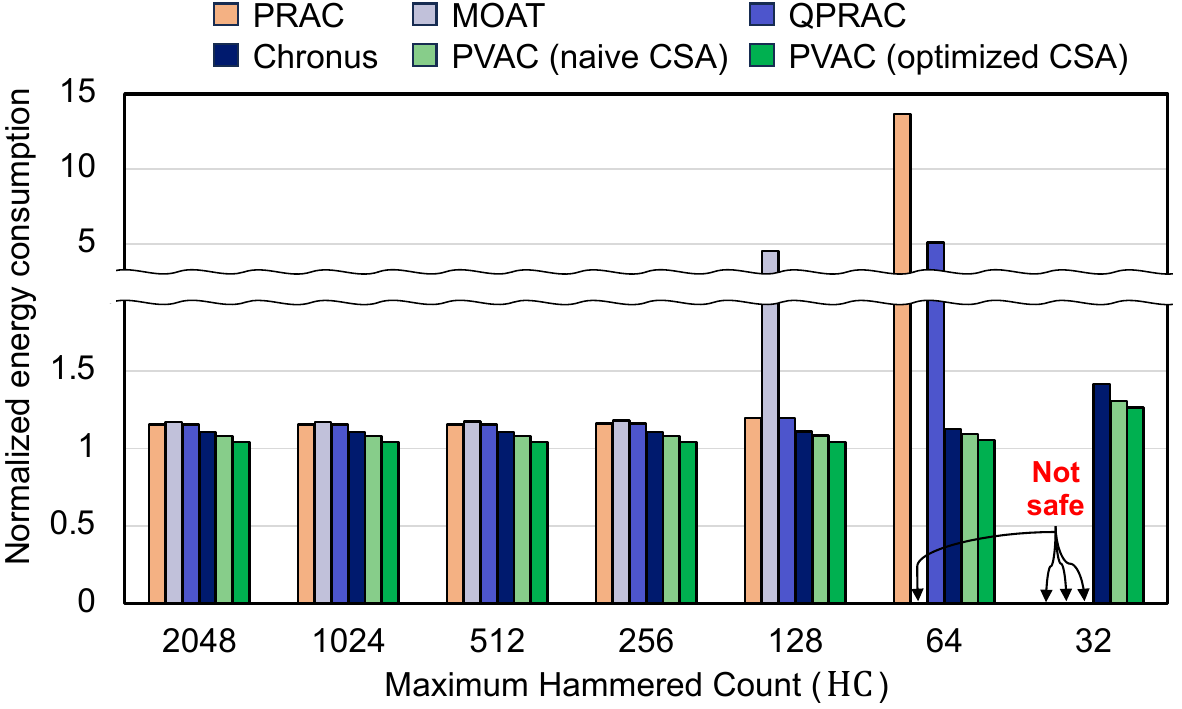}
  \vspace{-0.1in}
  \caption{Normalized average energy consumption across different workloads with different HC values.}
  \label{fig:benign_energy}
  \vspace{-0.1in}
\end{figure}

\noindent
\textbf{Energy.}
We evaluate the mean energy consumption of each mitigation scheme across all workloads (Figure~\ref{fig:benign_energy}).
Following the methodology of prior work~\cite{isca-2013-charm, hpca-2017-soup,isca-2020-clrdram,isca-2019-crow,hpca-2025-chronus}, we estimate CSA energy consumption during both normal accesses and \refcmd operations using SPICE simulations.

The optimized CSA design of PVAC reduces the number of activations during normal \refcmd operations, thereby achieving lower energy overhead than the na\"ive CSA design. 
Because the na\"ive CSA may activate up to eight rows per \refcmd (\S\ref{subsec:5_3_energy_overhead}), the energy overhead during an \refcmd operation can reach 161\% of normal access energy.
In contrast, the optimized
CSA design requires only a single CSA activation, resulting in an additional energy overhead of only 19.3\%.

During normal memory accesses, the na\"ive CSA increases the normal access energy by 20.1\%, while the optimized CSA increases it by 19.8\%.
This lower energy consumption of the optimized CSA is obtained even though the dual-CSA activation is required with a probability of \sfrac{3}{128}.
It is because the optimized CSA halves the number of rows per CSA, thereby reducing the energy consumption per CSA access compared to the naive CSA design.

Overall, the optimized CSA exhibits lower energy consumption than the na\"ive design.
This optimization reduces total energy consumption by 3.9\% compared to the na\"ive design across all HC values, primarily by lowering energy overhead during normal \refcmd operations.

PVAC and Chronus is more energy efficient than the other schemes, despite the additional energy overhead in the CSA.
As DRAM energy consumption is proportional to operation latency, timing parameters are the dominant factor. 
Although \actcmd and \precmd in PVAC and Chronus incur nearly a 20\% per-access energy increase due to CSA, this cost is less than the increase caused by PRAC timing parameters, consistent with the findings of Chronus~\cite{hpca-2025-chronus}.
When HC is 2048, even though \alerts are rarely triggered, PRAC, MOAT, and QPRAC incur energy overheads of 15.4\%, 17.2\% and 15.4\%, respectively, whereas PVAC and Chronus incur only 4.08\% and 10.8\%, respectively, despite performing proactive mitigations.

PVAC consumes less energy than Chronus as it performs fewer proactive mitigations.
Chronus triggers proactive mitigation once every two \trefi intervals, whereas PVAC initiates proactive mitigation only when a counter exceeds its threshold (\ie, \nbo/2).
Thus, PVAC avoids unnecessary proactive mitigations, lowering overall energy consumption.
Even if Chronus adopts the same proactive mitigation threshold, PVAC still performs fewer mitigations due to its larger security-equivalent \nbo, thereby reducing the mitigation energy overhead.

When the HC values drop below 256, Chronus issues more \rfmcmd commands than PVAC, which further increases its energy consumption. 
When HC is greater than 128, the energy overheads of Chronus and PVAC remain at 10.8\% and 4.08\%, respectively.
However, as HC decreases to 64, the energy overheads of Chronus and PVAC become 12.6\% and 5.35\%, respectively.
When HC is even reduced to 32, the overheads increase to 41.7\% and 26.4\%, respectively, with Chronus exhibiting the larger increase.

PRAC, QPRAC, and MOAT generally consume more energy than PVAC and Chronus.
MOAT incurs the highest energy overhead as its performance slowdown caused by the longer \trfc further increases refresh-related energy, and its restriction to \nmit = 1.
PRAC shows the next highest energy consumption due to the frequent \rfmcmd commands issued, while QPRAC reduces this overhead by proactive mitigations.

\subsection{Malicious attacks}
\label{subsec:7_4_Malicious_attack}

We evaluate the robustness of PVAC and Chronus under adversarial patterns with concurrent benign workloads.
We note that Appendix~\ref{sec:appendix} provides the theoretical upper bound of bandwidth degradation based on the methodology of Chronus~\cite{hpca-2025-chronus}, whereas this section focuses on the behavior observed under such workload conditions.

\noindent
\textbf{Adversarial pattern.}
To evaluate robustness against malicious attack patterns, we assume that one core issues adversarial memory accesses while the remaining three cores run benign workloads.
We focus on Chronus and PVAC as these schemes maintain competitive
performance overheads relative to other schemes under benign workloads (Figure~\ref{fig:benign_perf}).
The attacker activates $n$ rows in a round-robin manner to intentionally generate frequent \alerts and degrade overall system performance.
We vary the number of rows $n$ ($n \in \{8, 32, 128, 512, 1\mathrm{K}, 4\mathrm{K}, 8\mathrm{K}\}$) and sweep the maximum HC values from 32 to 512, because neither scheme triggers \alerts when the HC value exceeds 512.

\begin{figure}[!tb]
  \center
\includegraphics[width=0.98\columnwidth]{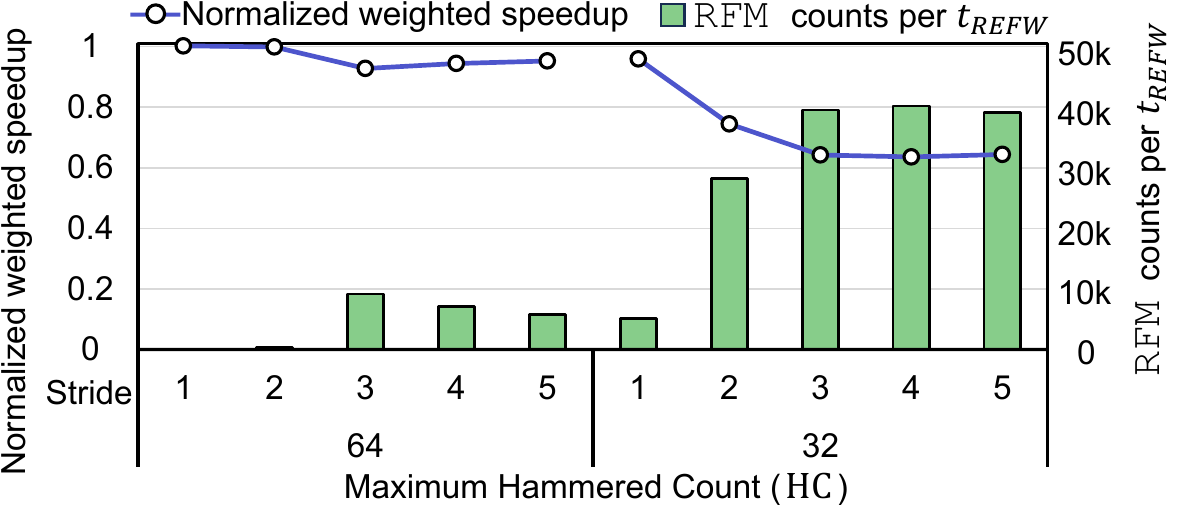}
  \vspace{-0.1in}
  \caption{Normalized weighted speedup across workloads under adversarial access patterns for varying maximum hammered count (HC) and row stride, with the number of \rfmcmd commands issued per \trefw.}
  \vspace{-0.06in}
  \label{fig:perf_attack_stride}
\end{figure}

We sweep the aggressor row stride from 1 to 5 to examine how the stride affects PVAC.
Because PVAC increments the counters of multiple victim rows, the aggressor row stride affects counter accumulation, which in turn changes the number of issued \rfmcmd commands.
Figure~\ref{fig:perf_attack_stride} shows the performance of PVAC, averaged across workload groups and $n$ values for each row stride, along with the number of \rfmcmd commands issued per \trefw.
No \alert is triggered when the HC values are 512, 256, and 128.
When the HC values decrease to 64 and 32, the largest \rfmcmd commands are issued at strides of 3 and 4, respectively.
As the performance degradation is proportional to the number of \rfmcmd commands issued, these cases are the most harmful among the adversarial patterns we evaluated.
When the stride is small (\eg, stride = 1 or 2), the counter increment can be immediately reset by the activation of the neighboring row, resulting in fewer \alerts.
For strides of 3 or larger, this immediate counter reset disappears, allowing disturbance to accumulate more effectively.
We derive an adversarial access pattern based on this spatial behavior to maximize issuing \rfmcmd commands.
We observe a significant increase in the number of issued \rfmcmd commands within \trefw when the stride is 3 or larger.
Thus, a stride of 3 represents the physical worst-case access pattern that avoids self-resetting penalties.

\begin{figure}[!tb]
  \center
  \includegraphics[width=0.98\columnwidth]{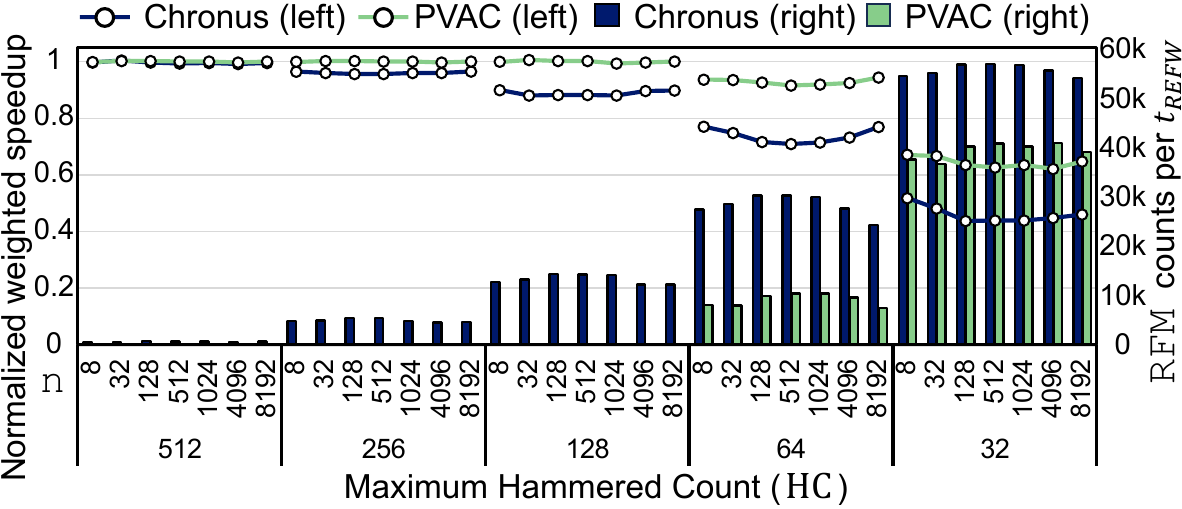}
  \vspace{-0.1in}
  \caption{Normalized weighted speedup across workloads under adversarial access patterns for varying maximum hammered counts (HC) and row pool ($n$), with the number of \rfmcmd commands issued per \trefw.}
  \vspace{-0.02in}
  \label{fig:perf_attack_rowpool}
\end{figure}

\noindent
\textbf{Performance.}
Figure~\ref{fig:perf_attack_rowpool} evaluates performance under the previously identified stride while varying $n$.
Due to 1) the larger \nbo and 2) the counter reset during normal refresh operations, PVAC achieves higher performance than Chronus across the evaluated settings.
In particular, when HC is 64 and $n = 128$, PVAC outperforms Chronus up to 29.4\%, which is the largest gap observed in our evaluation.

At small $n$ and HC values where \alerts can be triggered within \trefw, this difference is primarily associated with the larger \nbo of PVAC compared to Chronus.
As $n$ increases such that an \alert cannot be triggered within a single \trefw, PVAC still avoids triggering \alert due to its periodic counter reset on \refcmd, while Chronus continues to trigger \alerts.
For example, when HC = 512 and $n$ = 8K, the access pattern can persist across multiple \trefw intervals. 
Without refresh-driven counter reset, Chronus continues issuing \rfmcmd commands, reaching roughly 1K \rfmcmd commands.

\subsection{Sensitivity study}
\label{subsec:7_6_Sensitivity_Study}

Because PVAC updates the counters of both aggressor and victim rows, PVAC's feasibility highly depends on the counter update latency.
As explained in \S\ref{sec:5_pvac}, the counter update latency is determined by 1) the CSA timing parameters (\ie, \trcdcsa, \twrcsa, and \trpcsa) and 2) the number of counter values updated.
The CSA timing parameters depend on the number of CSA rows, which scales with the number of DSA rows~\cite{hpca-2017-soup,isca-2013-charm,tiered-2013-HPCA}, while the number of counters to update is determined by \br.
Thus, we analyze the counter update latency as a function of the number of rows per bank and \br.

\begin{figure}[!tb]
  \center
  \includegraphics[width=0.94\columnwidth]{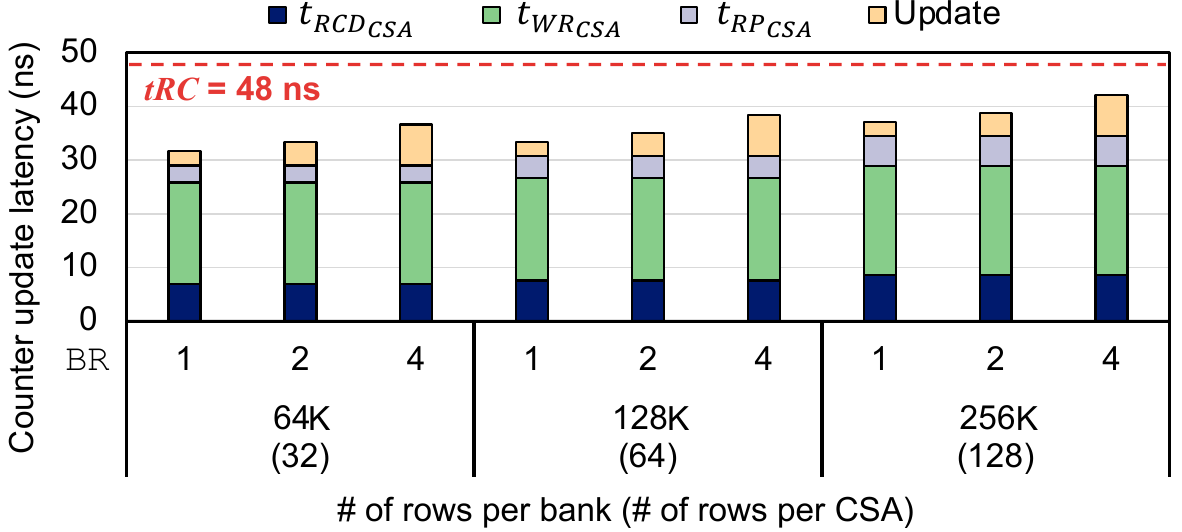}
  \vspace{-0.1in}
  \caption{Counter update latency under different row numbers and \br.}
  \label{fig:sensitivity}
  \vspace{-0.07in}
\end{figure}

Figure~\ref{fig:sensitivity} breaks down the latency into \trcdcsa, \twrcsa, \trpcsa, and \texttt{Update}, and shows that all configurations satisfy the \trc bound.
The \texttt{Update} component represents the latency to update the counters of both the victim and aggressor rows, computed as (2 $\times$ \br + 1) $\times$ \tup.
We evaluate three row counts specified by JEDEC~\cite{jedec-2024-ddr5}---64K, 128K, and 256K---and sweep \br up to 4, the maximum value defined by the standard, to calculate the latency.
All the evaluated configurations result in latencies below the 48~ns \trc constraint.

Increasing \br has a larger impact on the latency than increasing the row count.
When the number of rows increases from 64K to 128K and 256K at \br = 1, the CSA timing parameters account for 91.6\%, 92.0\%, and 92.8\% of the total latency, respectively.
In contrast, with 64K rows, the fraction of latency contributed by \texttt{Update} increases more significantly---from 8.39\% to 12.9\% and 20.8\%---as \br grows.
These results indicate that increases in \br amplify the counter update latency more substantially than scaling the row count. 

\section{Discussion}
\label{sec:8_discussion}

\noindent
\textbf{Distributed counter.}
PVAC introduces higher overhead on the counter structure compared to other schemes, which can be mitigated by distributing the counters across multiple chips.
In the current design, all DRAM chips maintain identical counter subarrays and perform update/reset operations in tandem.
By distributing the counters, each chip maintains counters for a subset of rows.
Then, the chips no longer maintain the same counters or perform identical operations.
Each only updates its dedicated rows, reducing counter maintenance overhead.

However, this approach requires an additional mechanism to broadcast the row index to be refreshed to all chips.
When an \alert is triggered in a specific chip, only that chip knows which row must be refreshed.
Therefore, the corresponding chip or MC must broadcast the row information to all chips within the \taboact window (\ie, before \rfmcmd is issued).

\noindent
\textbf{RowPress~\cite{isca-2023-rowpress}, ColumnDisturb~\cite{micro-2025-coldist}.}
RowPress is orthogonal to PRAC and can be effectively mitigated by the MC's page policy.
RowPress refers to a phenomenon where keeping a row activated for a long time induces bitflips in adjacent rows.
Modern CPUs primarily employ adaptive or close page policies~\cite{cal-2025-ohprac,lenovo-2023-pagepolicy,micro-2011-minimalist}.
An adaptive page policy performs \precmd after serving a certain number of \rdcmd/\wrcmd operations per \actcmd, while a close page policy performs \precmd immediately after a \rdcmd/\wrcmd when the MC does not have a pending request to the activated page.
With the limited row activation time, these policies naturally suppress RowPress-induced bitflips.

ColumnDisturb exhibits a behavior similar to \rh, where repeated activations in a subarray cause bitflips in neighboring subarrays, potentially affecting up to 3,072 rows.
Such a wide disturbance range makes conventional per-row counting mechanisms impractical due to its extremely large \br, thereby requiring a more coarse-grained counting approach.

Recent work proposes a subarray-level counting mechanism, SALT~\cite{salt-2026-HPCA}.
Instead of tracking activations at row granularity, SALT tracks the aggregate activation count of each subarray and refreshes rows at the subarray-level. 
To protect against ColumnDisturb, SALT increments the counter values of three subarrays---the given and its neighboring subarrays---thereby capturing disturbance propagation without incurring the prohibitive overhead of per-row tracking.
Because this work is built upon the PRAC protocol and does not depend on whether counting is performed on aggressors or victims, this mechanism can be adopted to PVAC with minimal modification.

\noindent
\textbf{Refresh postponing.}
Under PRAC, a maximum of two and four \refabcmd can be postponed in normal and fine granularity refresh mode, respectively~\cite{jedec-2024-ddr5}.
As PVAC resets counter values during periodic refresh operations, postponing a refresh also delays the corresponding reset.
With DDR5 \trc = 48\,ns, up to approximately 243 additional \actcmd commands can be issued to a single bank.
Consequently, additional \actcmd commands may accumulate before the reset occurs, potentially increasing the likelihood of triggering \rfmcmd and thereby incurring additional performance and energy overhead.
However, this is not unique to PVAC;
other PRAC-based schemes are equally exposed to the same extra number of \actcmd commands. 

\section{Related Work}
\label{sec:9_related_work}
\noindent
\textbf{Victim-based counting method.}
While several prior works explore victim-based counting schemes in \rh mitigation~\cite{dac-2019-mrloc, dac-2017-prohit, divide-tc-2025, sp-2022-protrr}, none of them directly address the fundamental limitations of aggressor-based counting methods.  
ProHIT~\cite{dac-2017-prohit} tracks potential victim rows using probabilistically managed hot and cold tables, effectively extending PARA~\cite{isca-2014-para}-like probabilistic refresh.
MRLoc~\cite{dac-2019-mrloc} also targets victim rows, but maintains them in a queue that derives a locality–aware weight so that frequently hammered victims receive higher refresh probability.

DIVIDE~\cite{divide-tc-2025} leverages in-DRAM caching to isolate frequently activated rows from neighboring rows, thereby reducing effective hammering.
It employs a lightweight victim-oriented mechanism in the cache tag table to track row vulnerability within the cache array.
ProTRR~\cite{sp-2022-protrr} builds upon TRRideal, an idealized victim-based counting model that assumes the ability to track the activation count of every potential victim row.
Because maintaining exact per-row counts using SRAM incurs prohibitive area and energy overhead, ProTRR replaces exact counting with a frequent-item approximation that identifies the most frequently activated rows using a Misra-Gries based algorithm~\cite{scp-1982-misragries, micro-2020-graphene}.

\noindent
\textbf{Side/covert-channel in PRAC.}
The ABO protocol of PRAC has recently been shown to be susceptible to timing-based side and covert channels~\cite{micro-2025-pracsidechannel, isca-2025-pracbackfire}. 
This vulnerability arises because, once a row's activation count reaches \nbo, the subsequent issuance of \rfmcmd commands introduces observable timing differences in system-wide memory access latency.

To mitigate this, prior works have proposed two main countermeasures. 
First, they adopt timing-based \rfmcmd schemes that issue the \rfmcmd command at a fixed and periodic interval.
For example, TPRAC~\cite{isca-2025-pracbackfire} and FR-RFM~\cite{micro-2025-pracsidechannel} propose issuing \rfmcmd per fixed timing interval.
Second, RIAC~\cite{micro-2025-pracsidechannel} suggests initializing the counters with a random value to make the exact moment of \rfmcmd issuance unpredictable.
During system initialization, each activation counter is assigned a randomly chosen non-zero value.

Both of them can be seamlessly adopted in PVAC to defend against the timing channels. 
Further, PVAC enhances security by resetting its counter upon commands that include activations (\eg, \actcmd, \refcmd,  and \rfmcmd), which fundamentally limits the overall frequency of \rfmcmd.
Therefore, PVAC is capable of providing a defense against timing channels by increasing the unpredictability of the mitigation process.
\section{Conclusion}
\label{sec:10_conclusion}

We have proposed PVAC, the first PRAC-based victim counting architecture that resolves the inherent limitations of aggressor-based PRAC mechanisms. 
Aggressor-based counting fails to account for incidental victim row activations or refreshes, overestimating disturbance accumulation.
In contrast, PVAC tracks the actual disturbance in a victim row, accurately measuring how many times it was attacked by the aggressor rows.
We embody PVAC's victim-based counting by introducing a counter subarray that allows concurrent updates to neighboring row counters, and optimize its organization so that PVAC operates under the default DDR5 timing parameters.
PVAC eliminates unnecessary counter accumulation, significantly reduces false \alerts, and consistently 
outperforms prior PRAC-based RowHammer mitigation solutions in both performance and energy. 
%


\section*{Acknowledgment}
This research was in part supported by Institute of Information \& communications Technology Planning \& Evaluation (IITP) grant funded by the Korea government (MSIT) [RS-2021-II211343, RS-2024-00402898], and
by the National Research Foundation of Korea (NRF) grant funded by MSIT [RS-2024-00405857]. 
The EDA tool was supported by the IC Design Education Center (IDEC), Korea.
This work was done when Minbok Wi was at Seoul National University (SNU).
Jung Ho Ahn, the corresponding author, is with the Department of Intelligence and Information and the Interdisciplinary Program in Artificial Intelligence, SNU.

\balance
\bibliographystyle{IEEEtranS}
\bibliography{refs}

\section{Appendix}
\label{sec:appendix}
We analyze the theoretical maximum performance degradation under an adversarial pattern, following the methodology of Chronus~\cite{hpca-2025-chronus}.
An adversarial access pattern can exploit the back-off mechanism to intentionally trigger frequent \rfmcmd commands and degrade performance.
To quantify the resulting upper bound, we adopt an analytical model that calculates the maximum fraction of bandwidth consumed by \rfmcmd as follows:
\begin{equation}
\vspace{-0.2em}
\label{eq:bw_degradation}
\Scale[0.85]{
\begin{split}
\frac{N_{Mit} \times t{ABO_{Recovery}}}{(N_{Mit} \times t{ABO_{Recovery}}) + (N_{BO} \times t_{RC})}
\end{split}}
\vspace{-0.15em}
\end{equation}
It captures the minimum time required to trigger one back-off event (\nbo$\times$\trc) and the subsequent blocking window caused by mitigative actions (\nmit$\times$\taborec).
Chronus derives this bound using a single-row stream as repeatedly activating one row at the minimum \trc interval maximizes the rate at which a single counter reaches \nbo.
This derivation remains directly applicable to PVAC because it relies only on two generic properties: 1) the counter increases once per \trc cycle and 2) reaching \nbo deterministically triggers the ABO protocol.
These two conditions also hold in PVAC for a victim-row counter; thus, repeatedly driving a PVAC victim counter to \nbo also maximizes the mitigation rate.
The same analysis directly characterizes the mitigation-induced bandwidth loss of PVAC under an idealized adversarial stream.

Each \alert in PRAC and PVAC incurs the fixed mitigation cost (\eg, \nmit = 4 for PRAC-4 and PVAC-4).
PVAC can be configured with a larger \nbo than PRAC (\S\ref{sec:6_security_analysis}); thus, the time required to trigger a back-off event of PVAC increases, reducing the frequency of mitigative actions.
For example, when HC is 256, \nbo of PVAC-4 can be set to 237, while PRAC-4 is restricted to 52.
Thus, the theoretical maximum bandwidth consumption for PVAC-4 is only 11.0\%, compared to 34.1\% for PRAC-4.
Because \nmit is predefined as 1, 2, or 4 for both PRAC and PVAC, this gap in bandwidth consumption remains consistent across all \nmit configurations.

Chronus adapts the mitigation cost to the actual number of aggressor rows whose counter values exceed \nbo, potentially reducing the mitigation overhead.
Although PVAC can be configured with a larger \nbo, Chronus may still incur lower mitigation cost because it issues \rfmcmd commands adaptively only for the aggressor rows that actually exceed \nbo within a single \alert.
Thus, when only one row reaches \nbo, Chronus performs only one \rfmcmd, effectively behaving as if \nmit were 1, whereas PVAC still incurs up to four \rfmcmd commands.
For example, when HC is 64, the theoretical maximum bandwidth consumption of PVAC-1 is 28.8\%, whereas Chronus exhibits 32.7\% under the same effective \nmit of 1.
In contrast, PVAC-4 uses a fixed \nmit of 4, which leads to a theoretical bandwidth consumption of 40.4\% despite its larger \nbo.
Thus, Chronus is more favorable than PVAC-4 in the theoretical single-row analysis.
In practical settings, however, interactions with benign workloads and other mitigation mechanisms can lead to behavior that differs from the analytical result.
We evaluate these effects in \S\ref{sec:7_evaluation}.

\end{document}